\documentclass[twocolumn,final,aps,prb,showpacs,superscriptaddress,amsmath,amssymb,amsfonts,floatfix]{revtex4-2}
\usepackage{epsfig}
\usepackage[dvipsnames,usenames]{color}
\usepackage{color}
\usepackage{amsmath}
\usepackage{comment}
\usepackage{array}
\usepackage{multirow}
\usepackage{tabularx}
\usepackage{float}
\usepackage[colorlinks=true,bookmarks=true, pdfborder={0 0 0},linkcolor=blue,urlcolor=blue,     breaklinks=true,bookmarksnumbered=true]{hyperref}

\begin{document}
\title{Hybridization induced quantum phase transition in bilayer Hubbard model}
\author{Xun Liu}
\affiliation{School of Physical Science and Technology, Soochow University, Suzhou 215006, China}

\author{Mi Jiang}
%\email{jiangmi@suda.edu.cn}
\affiliation{School of Physical Science and Technology, Soochow University, Suzhou 215006, China}
\affiliation{Jiangsu Key Laboratory of Frontier Material Physics and Devices, Soochow University, Suzhou 215006, China}

\begin{abstract}
Inspired by the recent experimental report on the pressure induced superconductor-insulator transition in cuprate superconductors as well as the superconductivity of the Ruddlesden-Popper-phase La$_3$Ni$_2$O$_7$ under high pressure, we systematically investigated the single-orbital bilayer Hubbard model in the regime of large interlayer hybridization to mimic the pressure effects. We map out the phase diagram of interlayer hybridization versus density in the regime of intermediate to strong hybridization. In particular, we found that the sufficiently strong hybridization can destroy the $s^{\pm}$-wave pairing and induces its transition to correlated metallic, pseudogap, and Fermi liquid phases depending on the doping regime. The phase diagram hosted by the bilayer model implies its role as the versatile platform to explore the pressure effects on the two-dimensional to three-dimensional crossover physics of Hubbard-type models.
\end{abstract}

\maketitle

\section{Introduction}
Recent experiment observed a pressure induced quantum phase transition from a superconducting state to an insulating-like state in Bi$_2$Sr$_2$CaCu$_2$O$_{8+\delta}$ (Bi2212) superconductors with two CuO$_2$ planes in a unit cell for a range of doping levels~\cite{Sunliling2022}, which is distinct from the more familiar transition between superconducting and metallic states by chemical doping~\cite{Keimer2015}.
Therefore, what kind of non-superconducting state and how this state develops out of superconductivity (SC) can be crucial for understanding the unconventional SC in cuprates.

Although the same phenomenon holds true for other cuprates with a single or three CuO$_2$ planes in a unit cell~\cite{Sunliling2022}, it naturally motivates for a theoretical exploration of bilayer single-orbital Hubbard model as the minimal model to capture the pressure effect normally mimicked by the interlayer hybridization.
% bilayer model (single-orbital)
As a versatile platform, the bilayer Hubbard model has been widely adopted to describe the strongly correlated electronic systems such as the antiferromagnetic ordering, spin fluctuation mediated pairing, and interlayer physics induced by the hybridization and/or interaction~\cite{bilayer1,bilayer2,bilayer3,bilayer4,bilayer5,bilayer6,bilayer7,bilayer8}. Some recent proposals regards on the enhancement of SC~\cite{incipient2,incipient3,bilayer5,Werner, Maier2022}. 
One significant ingredient of any bilayer models lies in the interlayer hybridization and/or interaction, which can be simply treated as mimicking the experimental pressure effects. 
Interestingly, recent high pressure experiments on single crystals of rare-earth nickelate La$_3$Ni$_2$O$_7$ discovered the SC with maximum $T_c$ of 80 K at pressures between 14.0 GPa and 43.5 GPa~\cite{327}. Despite of its essential multi-orbital nature, its bilayer lattice structure also points to the exploration of the bilayer Hubbard models.

In our recent investigation of bilayer multi-orbital model adopting the cluster exact diagonalization~\cite{Mi23}, we studied different parameter regimes potentially relevant to both cuprates and
the newly discovered infinite-layer nickelate SC~\cite{2019Nature,Aritareview,Botana_review,Held2022,Hanghuireview}. Some interesting physics were revealed in terms of the asymmetric electronic distribution between two layers induced by interlayer hybridization.
Unfortunately, in spite of its complete consideration of all $3d$ orbitals, the computational cost challenge impedes its full many-body treatment.

Hence, in this work, we focus on the simpler bilayer single-orbital Hubbard model aiming to examine the fate of the SC upon the strong interlayer hybridization for a wide range of hole doping levels. Specifically, we map out the phase diagram within this intermediate to strong hybridization regime. We mention that the bilayer model is not simply a toy model relevant to condensed matter physics but has been realized with ultracold atomic systems that observing the competing magnetic orders~\cite{bilayer4}. Therefore, our systematic exploration here can be potentially realized in the future cold atomic experiments.

%%%%%%%%%%%%%%%%%%%%%%%%%%%%%%%%%%%%%%%%%%%%%%%%%%%%%%%%%%%%%%%%%%%%%%%%%%%
\section{Model and Method}\label{model}
\subsection{Bilayer Hubbard model}
We consider the bilayer Hubbard model, defined on a two-dimensional square lattice
\begin{align}\label{eq:HM}
	\hat{H} = &- \sum_{ i,j,\alpha,\sigma} t_{ij}(c^\dagger_{i,\alpha,\sigma}c^{\phantom\dagger}_{j,\alpha,\sigma}+h.c.) -\mu\sum_{i,\alpha,\sigma}n_{i,\alpha,\sigma} \notag\\
 -  &V\sum_{i,\sigma}(c^\dagger_{i,1,\sigma}c^{\phantom\dagger}_{i,2,\sigma}+h.c.)\notag + U\sum_{i,\alpha}n_{i,\alpha,\uparrow}n_{i,\alpha,\downarrow}\notag
\end{align}
where $c^\dagger_{i,\alpha,\sigma}/c^{\phantom\dagger}_{i,\alpha,\sigma}$ creates/annihilates an electron with spin $\sigma$(=$\uparrow$, $\downarrow$) on the $\alpha$-th layer($\alpha$=1,2). The intralayer hopping term $t_{i,j}$ includes the nearest neighbor $t$ and the next-nearest neighbor hoppings $t'$. The interlayer hybridization $V$ and the on-site Coulomb interaction $U$ are two important parameters. The chemical potential $\mu$ controls the occupancy. Throughout this work, we set the nearest neighbor hopping $t=1$ as the unit of energy. Besides, the nearest neighbor hopping $t'=-0.15t$ and $U=7t$ are adopted. The non-interacting dispersion reads as
\begin{equation} \label{dis}
	\epsilon_k = -2t(\cos k_x+\cos k_y)-4 t' \cos k_x\cos k_y \\
        - V\cos k_z - \mu 
\end{equation}
where the combinations of bonding ($k_z=0$) and anti-bonding band ($k_z=\pi$) can be viewed as Bloch states of a one-band model in three-dimension (3D). In other words, we treat the bilayer model as the simplest 3D Hubbard model with only two layers~\cite{bilayer5}.

\subsection{Dynamical cluster approximation (DCA)} 
DCA with the continuous-time auxilary-field (CT-AUX) quantum Monte Carlo (QMC) cluster solver~\cite{Hettler98,Maier05,code,GullCTAUX} is employed to numerically solve the bilayer model.
As a celebrated quantum many-body numerical method, DCA evaluates the physical observables in the thermodynamic limit via mapping the bulk lattice problem onto a finite cluster embedded in a mean-field bath in a self-consistent manner~\cite{Hettler98,Maier05}, which is realized by the convergence between the cluster and coarse-grained (averaged over a patch of the Brillouin zone around a specific cluster momentum $\mathbf{K}$) single-particle Green's functions. 
In particular, the short-range interactions within the cluster are treated exactly with various numerical techniques, e.g. CT-AUX used in the present study; while the longer-ranged physics is approximated by a mean field hybridized with the cluster. Therefore, increasing the cluster size systematically approaches the exact result in the thermodynamic limit. 
The finite size of the cluster is essentially approximating the whole Brillouin zone by a discrete set of $\mathbf{K}$ points so that the self-energy $\Sigma(\mathbf{K},i\omega_n)$ is a constant function within the patch around a particular $\mathbf{K}$ and a step function in the whole Brillouin zone.
Generically, the quantum embedding methods including DCA have better minus sign problem than the finite-size QMC simulations because of the hosting mean field.
We refer to Ref.~\cite{Maier05} for more discussions on DCA technique and its insight on the strongly correlated electronic systems.

Most of our calculations were conducted with $N_c=8$-site DCA cluster to incorporate the momentum space resolution including nodal $\mathbf{K}= (\pi/2,\pi/2)$ and antinodal $\mathbf{K} = (\pi,0)$ directions. Larger $N_c=12$ is also employed for checking some results qualitatively.
We remark that larger $N_c$ does not lead to significant change of our major results in the strong interlayer hybridization $V$ regime, whose dominant local physics results in that a small $N_c$ should be sufficient to manifest the most essential physics discussed here in most situations.

The SC properties can be studied via solving the Bethe-Salpeter equation (BSE) in the eigen-equation form in the particle-particle channel~\cite{Maier2006,scalapino2007numerical}
\begin{align} \label{BSE}
    -\frac{T}{N_c}\sum_{K'}
	\Gamma^{pp}(K,K')
	\bar{\chi}_0^{pp}(K')\phi_\alpha(K') =\lambda_\alpha(T) \phi_\alpha(K)
\end{align}
where $\Gamma^{pp}(K,K')$ denotes the lattice irreducible particle-particle vertex of the effective cluster problem with combining the cluster momenta $\bf K$ and Matsubara frequencies $\omega_n=(2n+1)\pi T$ as $K=(\mathbf{K}, i\omega_n)$. 
The coarse-grained bare particle-particle susceptibility
\begin{align}\label{eq:chipp}
	\bar{\chi}^{pp}_0(K) = \frac{N_c}{N}\sum_{k'}G(K+k')G(-K-k')
\end{align}
is obtained via the dressed single-particle Green's function $G(k)\equiv G({\bf k},i\omega_n) =
[i\omega_n+\mu-\varepsilon_{\bf k}-\Sigma({\bf K},i\omega_n)]^{-1}$, where $\mathbf{k}$ belongs to the DCA patch surrounding the cluster momentum $\mathbf{K}$ with $\mu$ the chemical potential, $\varepsilon_{\bf k}=-2t(\cos k_x+\cos k_y)$ the
dispersion relation, and $\Sigma({\bf K},i\omega_n)$ the cluster self-energy. In practice, we choose 24 or more discrete points for both the positive and negative fermionic Matsubara frequency $\omega_n=(2n+1)\pi T$ mesh for measuring the four-point quantities like two-particle Green's functions and irreducible vertices. Therefore, the BSE Eq.~\eqref{BSE} reduces to an eigenvalue problem of a matrix of size $(48N_c)\times (48N_c)$.

The normal state pairing tendency is reflected by the leading eigenvalue $\lambda_\alpha(T)$ for pairing symmetry $\alpha$. Simultaneously, the associated eigenvector $\phi_\alpha(K)$ can be viewed as the normal state analog of the SC gap function~\cite{Maier2006,scalapino2007numerical}.
It has been widely accepted that the $d$-wave pairing plays a dominant role in the cuprate superconductors and closely relevant single-orbital Hubbard model~\cite{scalapino2007numerical,Hubreview1}. 
Nonetheless, the bilayer model hosts a transition from $d$-wave to $s^{\pm}$-wave with increasing interlayer hybridization $V$~\cite{bilayer8}. Therefore, instead of $d$-wave pairing, here we only focus on the $s^{\pm}$-wave pairing symmetry, where two electrons in opposite layers form Cooper pair with each other, induced by a relatively large $V/t \ge 2.0$.

%%%%%%%%%%%%%%%%%%%%%%%%%%
\section{Results}

\begin{figure}[t!]
\psfig{figure=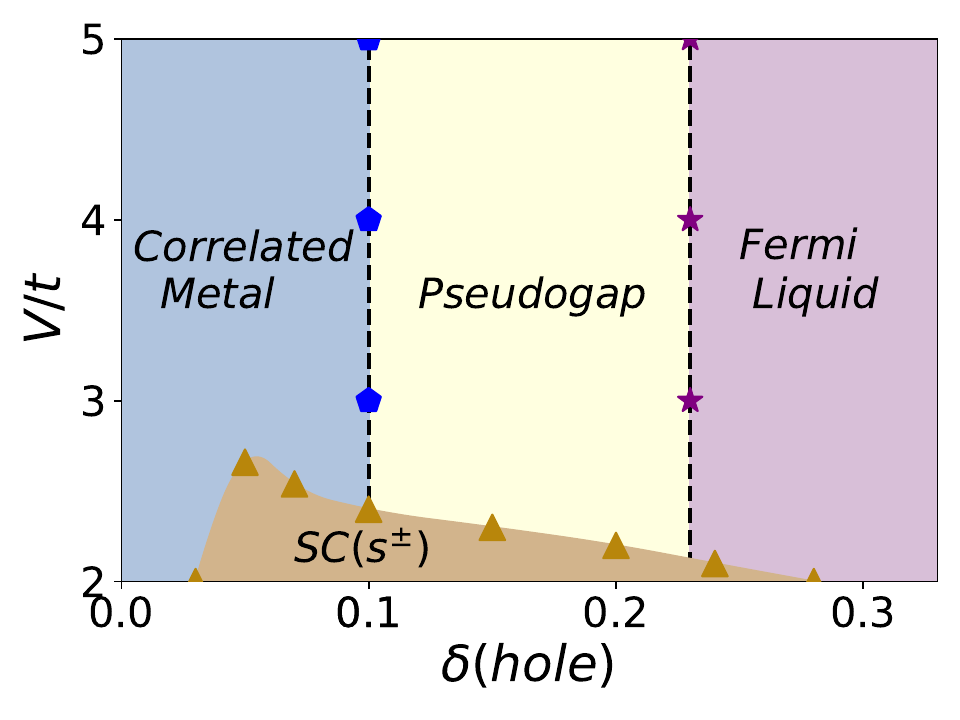,height=6.4cm,width=.49\textwidth, clip}
\caption{Tentative phase diagram of the bilayer Hubbard model with the intermediate hybridization $V/t$ and the hole doping level $\delta$ as control parameters.}
\label{phase}
\end{figure}

The major result of the present work is illustrated in the phase diagram Fig.~\ref{phase}, where the phase boundaries from the $s^{\pm}$-wave SC to various phases in different doping regime are estimated.
In the next sections, we will first focus on the $s^{\pm}$-wave pairing phase and then examine the fate of its transitions in different doping regimes.

\subsection{$s^{\pm}$ pairing: BSE eigenvalue}

\begin{figure}
\centering
\psfig{figure=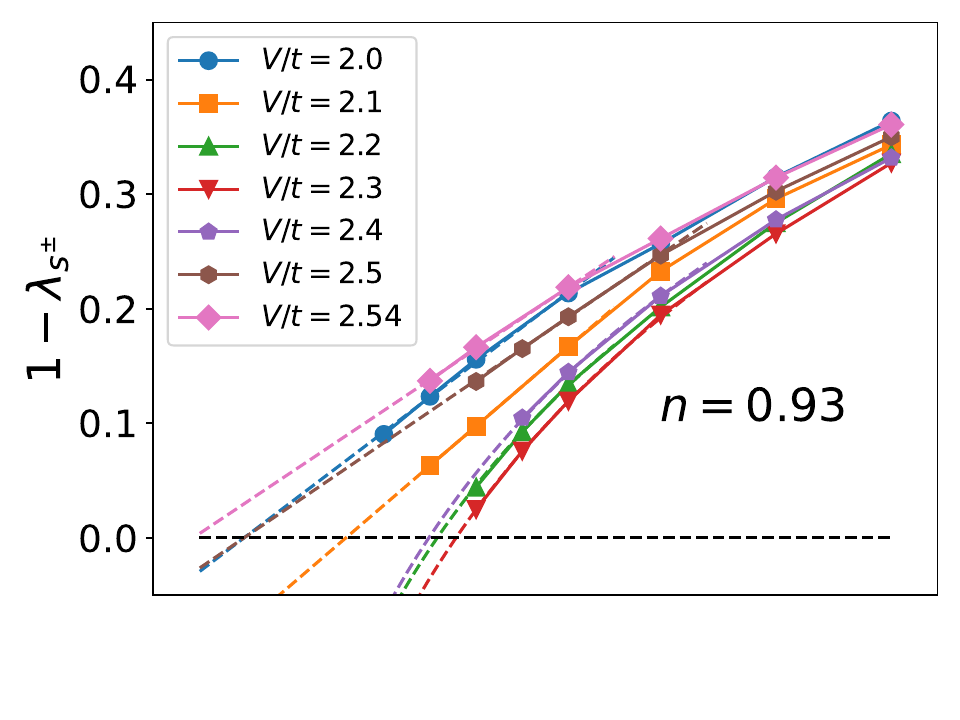,height=6.0cm,width=.49\textwidth, clip}

\vspace{-3em}

\psfig{figure=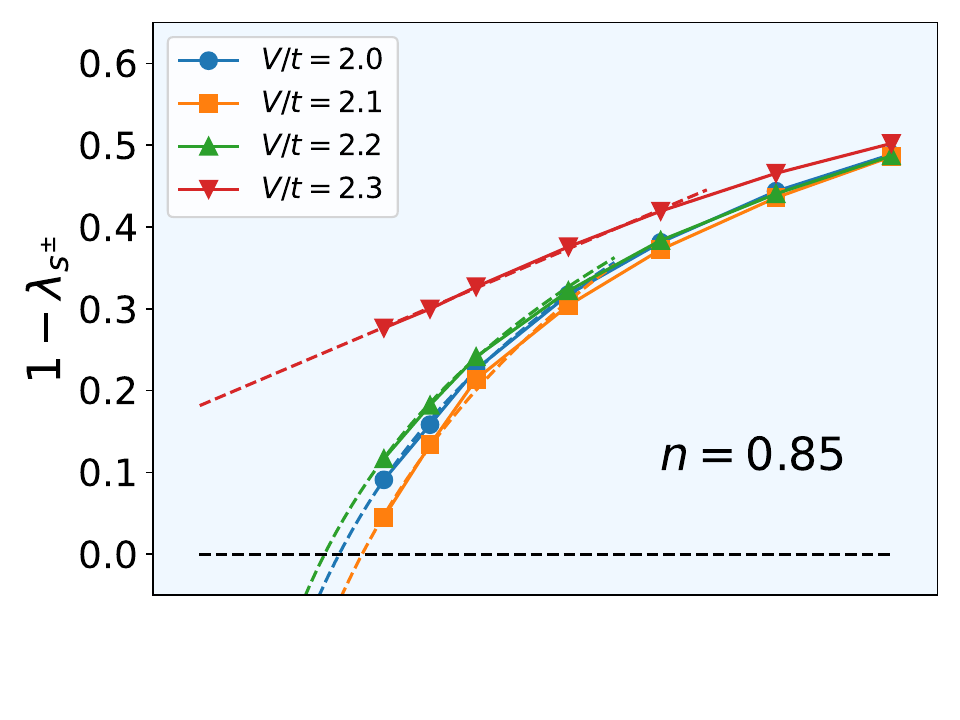,height=6.0cm,width=.49\textwidth, clip}

\vspace{-3em}

\psfig{figure=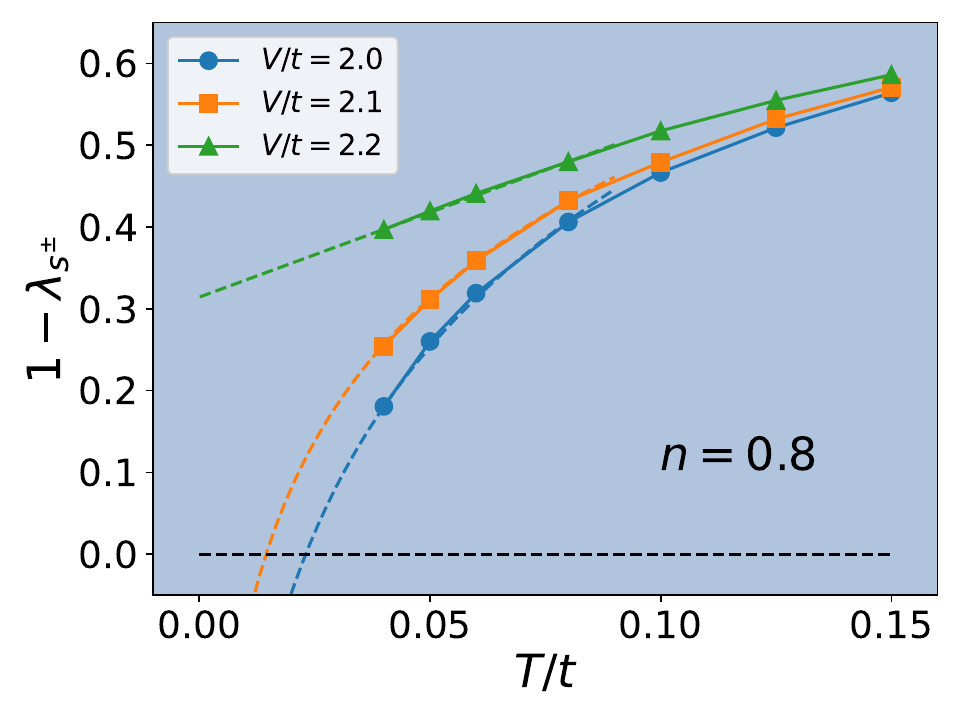,height=6.0cm,width=.49\textwidth, clip}
\caption{Temperature evolution of $1-\lambda_{s^\pm}(T)$ for varied hybridization $V$ at the characteristic fillings $n$=0.93, 0.85 and 0.8 of bilayer model.}
\label{ldspm}
\end{figure}

To investigate the influence of the $s^{\pm}$-wave SC properties, we first explored the temperature evolution of the leading $s^{\pm}$-wave BSE eigenvalue $\lambda_{s^{\pm}}(T)$ for varied doping. Fig.~\ref{ldspm} illustrates its behavior at three characteristic fillings $n=0.93, 0.85, 0.8$.

The weakened $s^{\pm}$-wave pairing tendency is manifested by the departure of $1-\lambda_{s^{\pm}}(T)$ curve from zero. Generically, from the magnitude of $1-\lambda_{s^{\pm}}(T)$, the pairing instability is strongest at larger density (lower doping) $n=0.93$ and diminishes with further doping for any fixed $V$.
Apparently, the impact of the hybridization is non-monotonic for all three typical dopings in that the intermediate $V$ pushes the curves closest to zero, indicating the promotion of $s^{\pm}$-wave pairing.
As expected, it is not surprising that larger dopings, for instance $n=0.8$, gradually destroy the pairing instability. Besides, large enough hybridization suppresses the SC with interlayer singlet formation of insulating nature.

One noticeable feature is that most curves exhibit BCS logarithmic temperature evolution, suggesting that these systems with relatively strong interlayer hybridization (before suppressing SC) are possibly dominated by the BCS Cooper pair fluctuations, which is reminiscent of the similar behavior observed in the $d$-wave SC of the conventional single-band Hubbard model in the overdoped regime~\cite{Maier2019}. For small or large $V$, the low temperature behavior shows the linear or even signals of exponential behavior, which implies the non-BCS type pair fluctuations as discussed for the pseudogap regime of the single-band model~\cite{Maier2019}. Hence, if the QMC sign problem, which is challenging to simulate low enough temperatures close to $T_c$, does not allow for the direct evaluation of the pairing $T_c$ via $1-\lambda_{s^{\pm}}(T_c)=0$, we have to estimate the $T_c$ by extrapolating these curves to lower temperatures by assuming that they obey the logarithmic or linear temperature evolution.

\begin{figure} 
\center
\psfig{figure=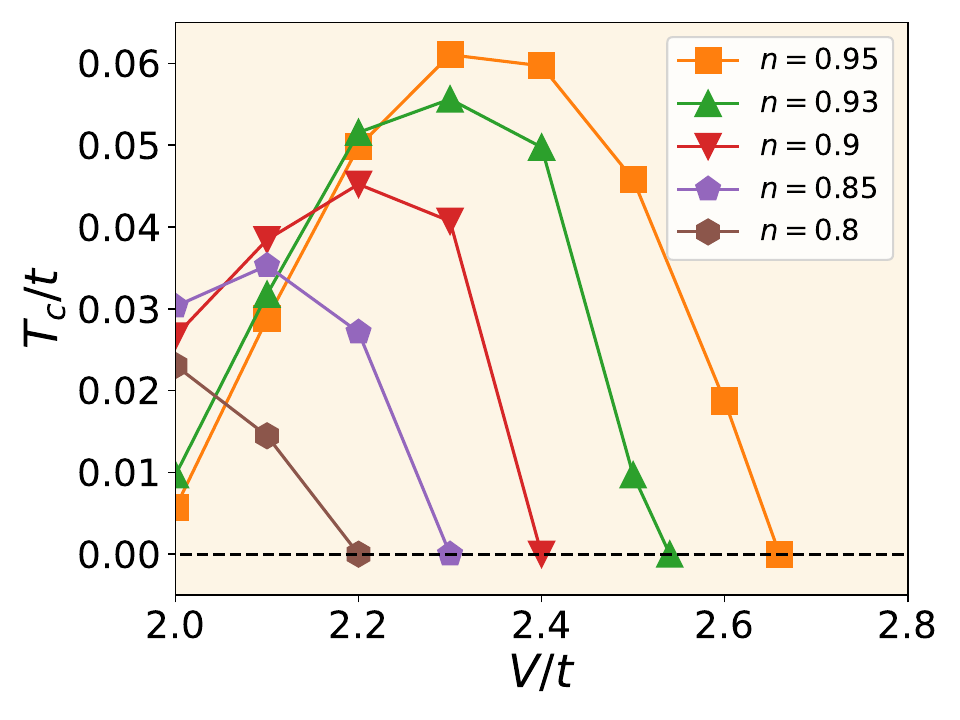,height=6.0cm,width=.49\textwidth, clip}
\caption{Interlayer hybridization dependence of the $s^{\pm}$-wave pairing transition temperature for different fillings.} 
\label{Tc}
\end{figure}

\begin{figure*}[t]
\centering
\psfig{figure=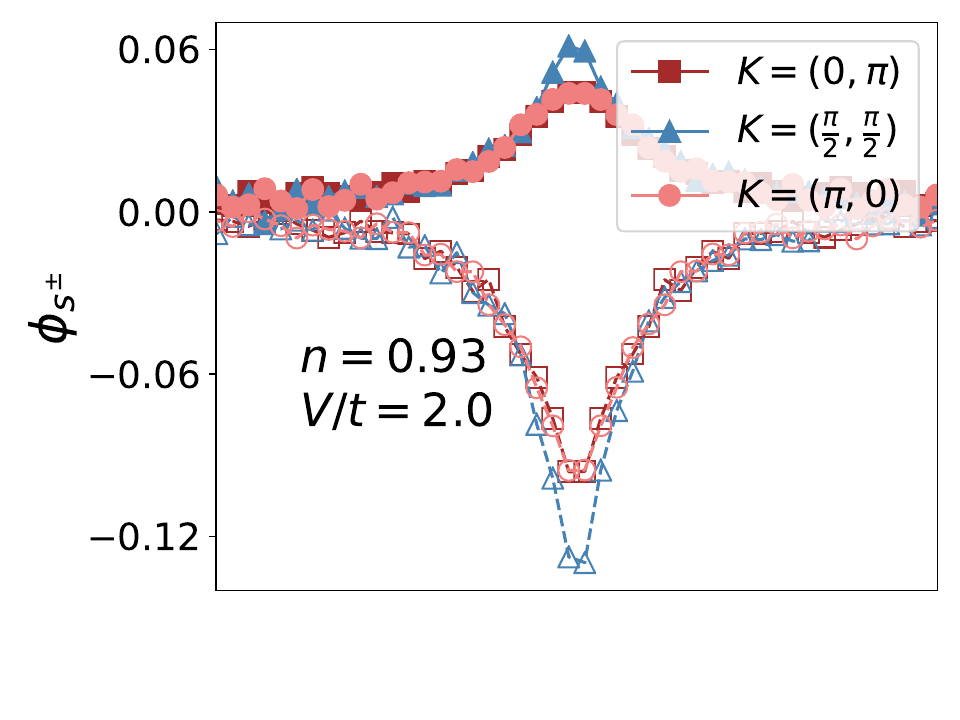,height=4.1cm,width=.32\textwidth, clip}
\vspace{-2em}
\hspace{-1em}
\psfig{figure=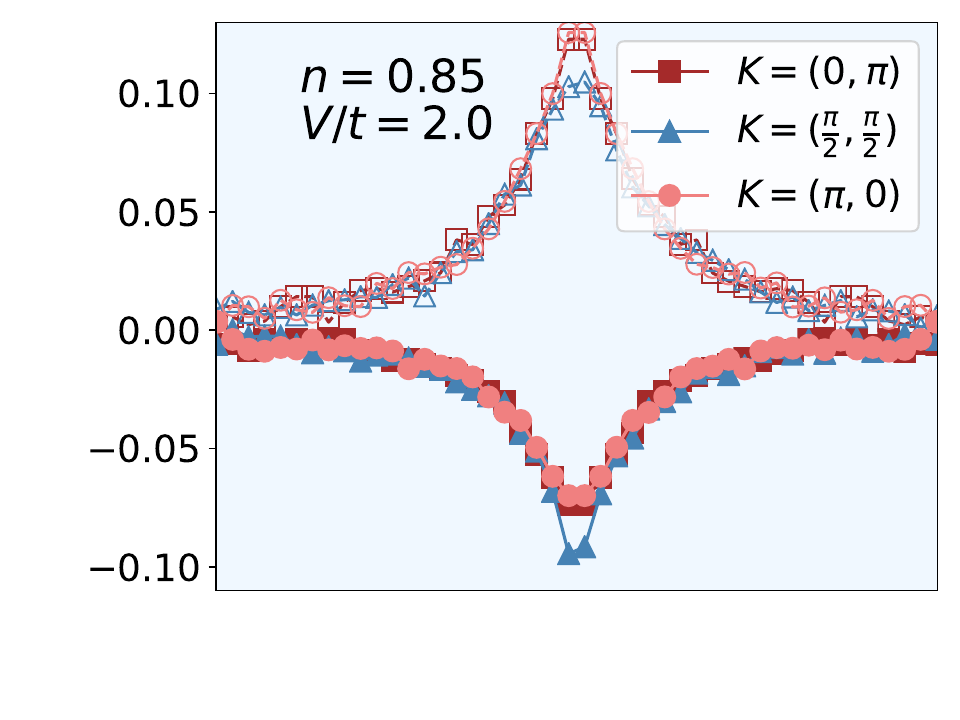,height=4.1cm,width=.32\textwidth, clip}
\hspace{-1em}
\psfig{figure=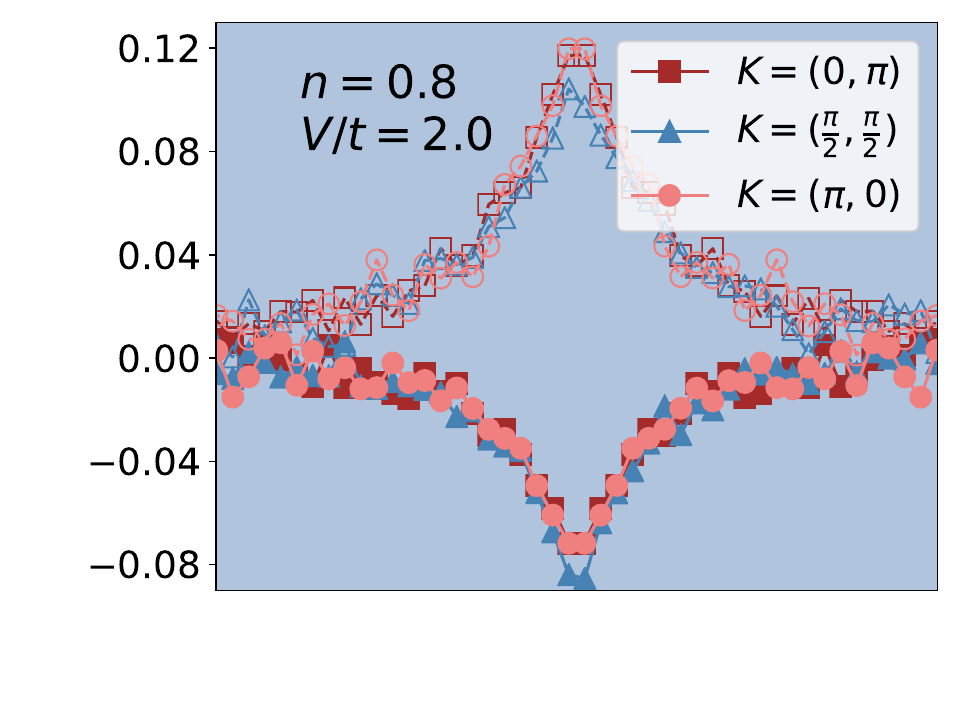,height=4.1cm,width=.32\textwidth, clip}
\psfig{figure=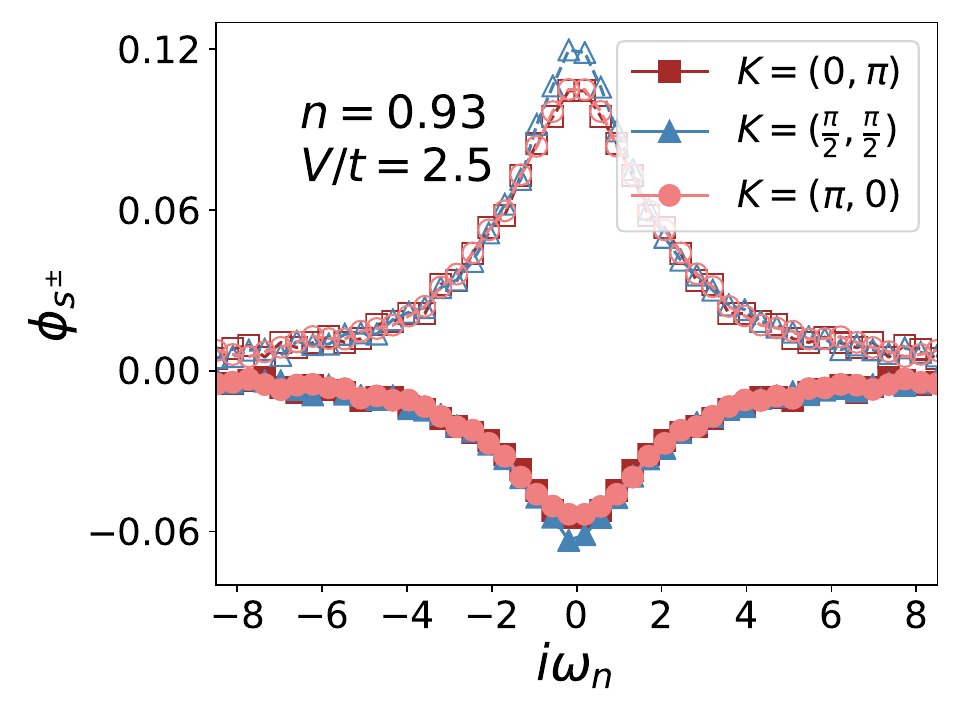,height=4.1cm,width=.32\textwidth, clip}
\hspace{-1em}
\psfig{figure=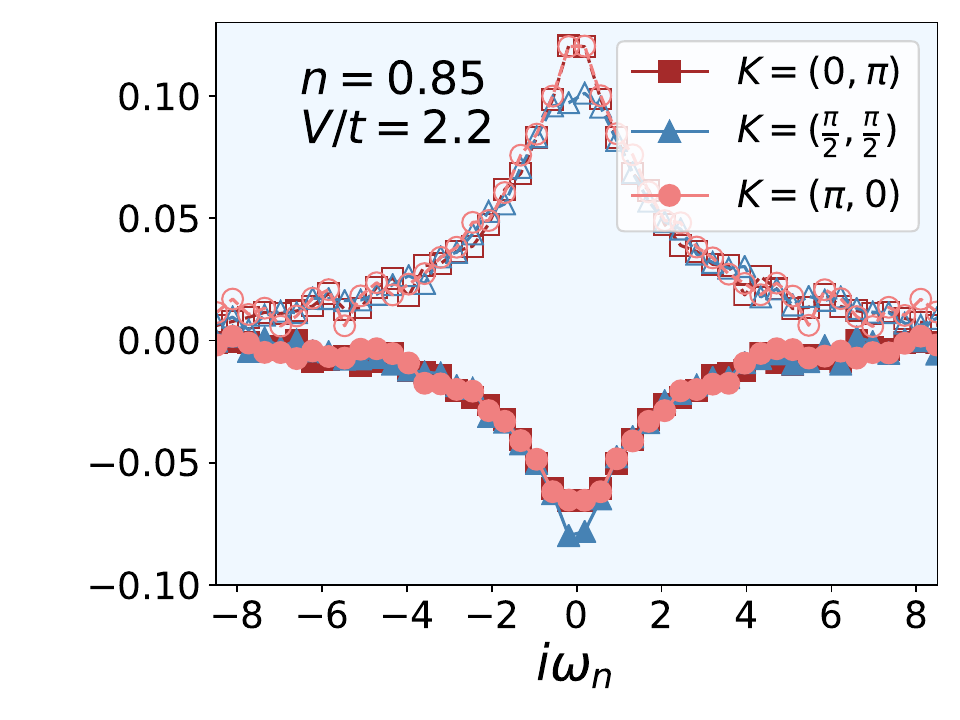,height=4.1cm,width=.32\textwidth, clip}
\hspace{-1em}
\psfig{figure=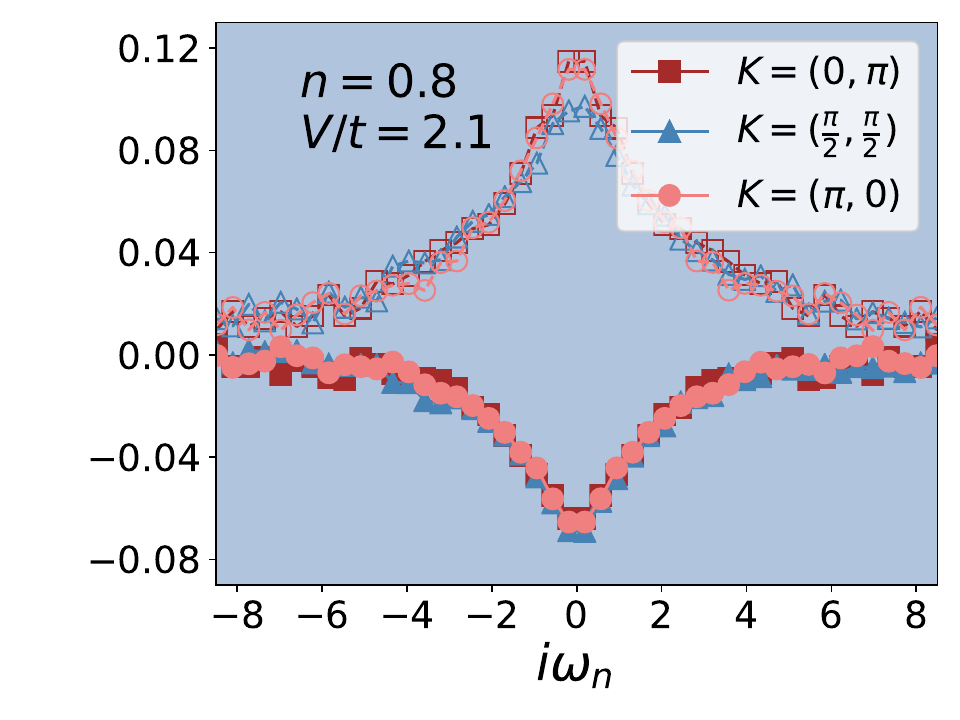,height=4.1cm,width=.32\textwidth, clip}
\caption{Frequency dependence of the leading eigenvector $\phi$ at various ${(\bf K_x, \bf K_y)}$ for $\bf K_z=0$ (solid lines) and $\bf K_z=\pi$ (dashed lines) with $T/t=0.06$ for $V$ within the superconducting regime of the phase diagram Fig.~\ref{phase}.}
\label{eigvec}
\end{figure*}

Fig.~\ref{Tc} reveals the non-monotonic dependence of the extracted $T_c$ with the magnitude of hybridization for a wide range of fillings. The generic feature lies in that intermediate $V$ is optimal for promoting the $s^{\pm}$-wave pairing while this instability is destroyed gradually by stronger hybridization, which facilitates the electronic structure reconstruction for an interlayer singlet formation. Apparently, the critical $V$ for suppressing SC decreases with the hole doping level, which matches with the phase diagram Fig.~\ref{phase}, where the critical doping $\sim 0.97$ for $V=2.0t$ is also estimated with the same procedure illustrated above (not shown) since there is no SC at half-filling as expected. Simultaneously, the optimal $V$ for highest $T_c$ decreases with the doping level as well. The fate of $s^{\pm}$ pairing after its disappearance is our next central topic discussed later.

%\textcolor{red}{The high pressure experiments~\cite{Sunliling2022} demonstrated some universal behavior of the renormalized $T_c$ versus the pressure. Here we performed the similar analysis by renormalizing the curves in panel (a) by the maximal $T_c$ and corresponding optimal hybridization $V_c$ for each filling $n$.
%Fig.~\ref{Tc}(b) shows the roughly universal shape of the detrimental impact of strong hybridization.}

\subsection{$s^{\pm}$ pairing: eigenfunction}

Fig.~\ref{eigvec} further illustrates the frequency dependence of the leading eigenvector $\phi_{s^{\pm}}$ at various ${(\bf K_x, \bf K_y)}$ for $\bf K_z=0$ (solid lines) and $\bf K_z=\pi$ (dashed lines).
For all the parameter sets, $\phi_{s^{\pm}}$ changes sign between $\bf K_z=0, \pi$ but
keeps the same sign as a function of in-plane ${(\bf K_x, \bf K_y)}$. In unconventional superconductors, the sign change in the gap function, or here our eigenfunction as its normal state analog, is to minimize the repulsion effects of the Hubbard interaction, which is well known as the $d$-wave pairing in cuprate SC and $s^{\pm}$ wave in the Fe-based SC. Fig.~\ref{eigvec} clearly shows that the in-plane  ${(\pi, 0)}$ and ${(0, \pi)}$ are exactly the same so that excludes the $d$-wave pairing symmetry, which is not surprising because of the relatively large interlayer hybridization while $d$-wave SC can only exist for the single layer limit when $V$ is relatively small~\cite{bilayer8}.

As shown in Fig.~\ref{eigvec}, for these three characteristic densities, the common feature lies in the strong anisotropy, namely strong ${(\bf K_x, \bf K_y)}$ dependence of $\phi_{s^{\pm}}$ at our smallest hybridization $V=2.0t$, gives way to the more isotropic behavior at larger $V$ within the SC regime of the phase diagram Fig.~\ref{phase}. Besides, this anisotropy-isotropy transition also occurs with  increasing the doping, as evidenced by the three columns from left to right for fixed $V$.
This observation can be intrinsically connected with the pseudogap (PG) phenomena in the single layer/band Hubbard model, where the strong momentum selective behavior dominates the system within the underdoped regime, e.g. $n=0.93$. We will discuss this point in the next section to provide more evidence on the remnant of this PG behavior in the strong hybridization limit.

\begin{figure*}[t]
\psfig{figure=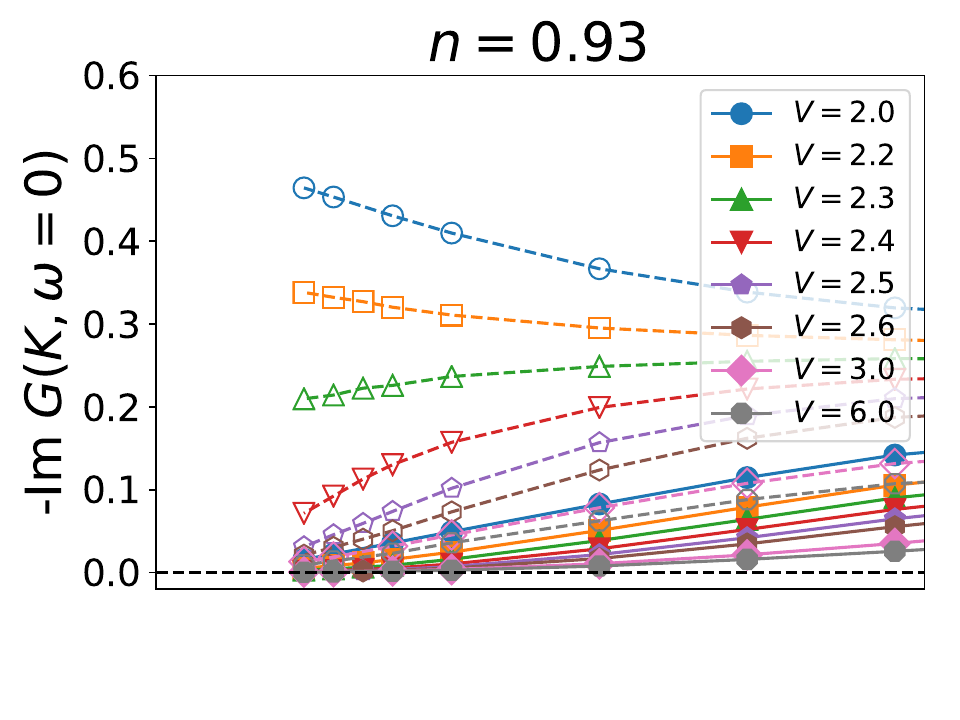,height=4.0cm,width=.32\textwidth, clip}
\vspace{-2em}
\hspace{-1em}
\psfig{figure=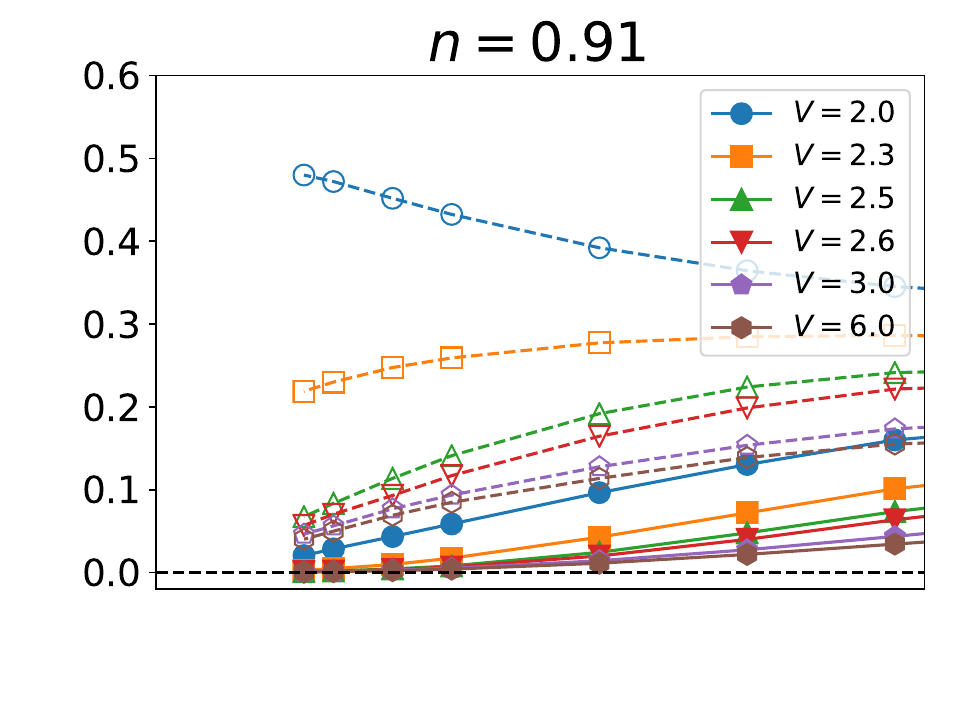,height=4.0cm,width=.32\textwidth, clip}
\hspace{-1em}
\psfig{figure=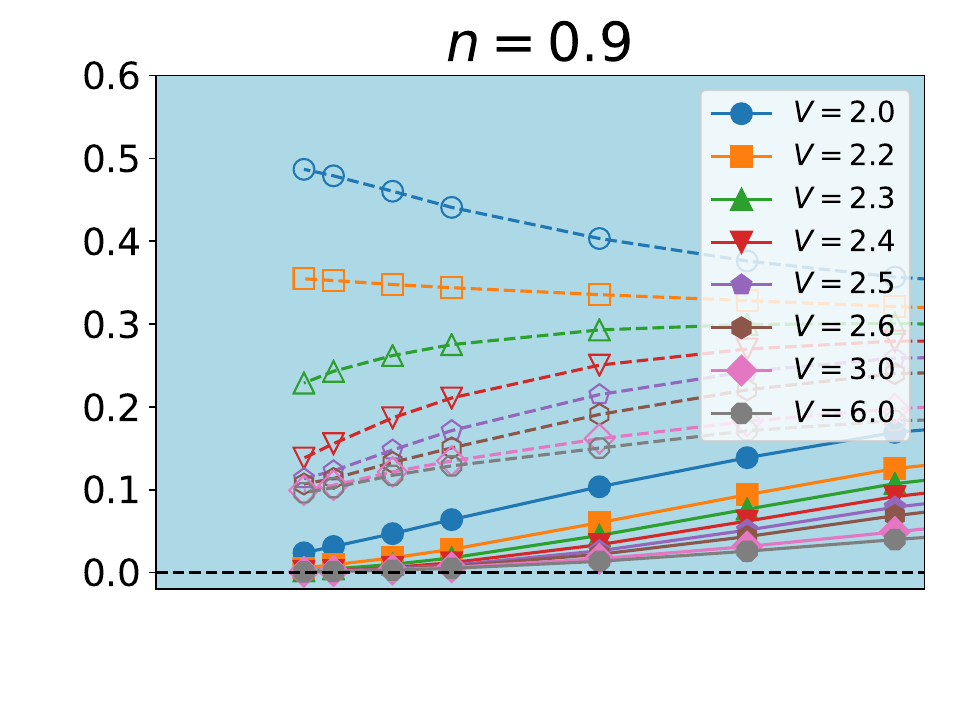,height=4.0cm,width=.32\textwidth, clip}
\psfig{figure=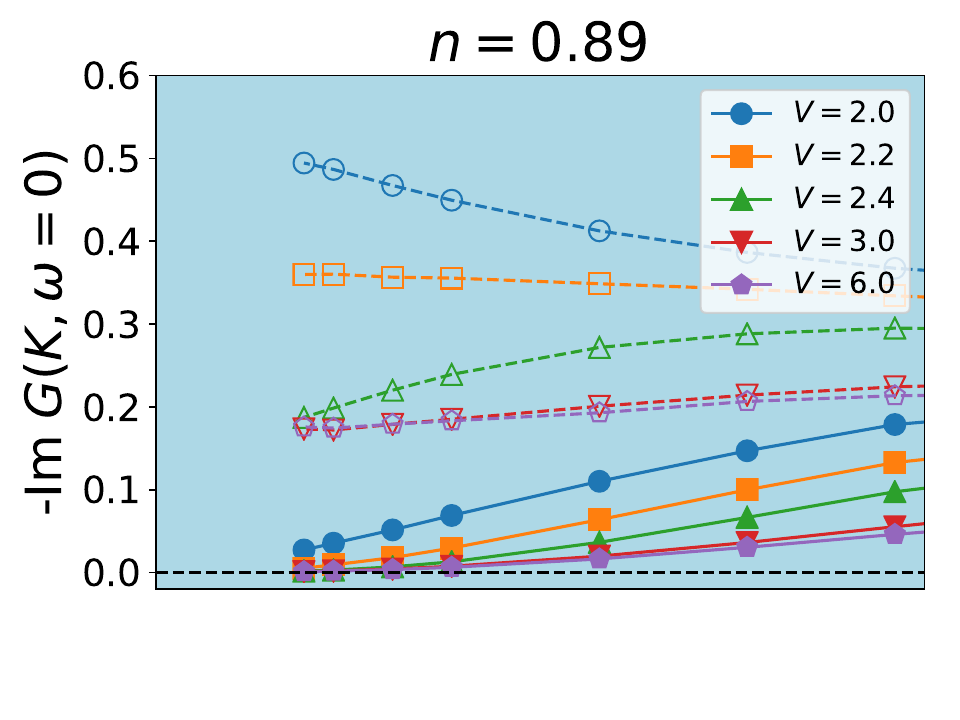,height=4.0cm,width=.32\textwidth, clip}
\vspace{-2em}
\hspace{-1em}
\psfig{figure=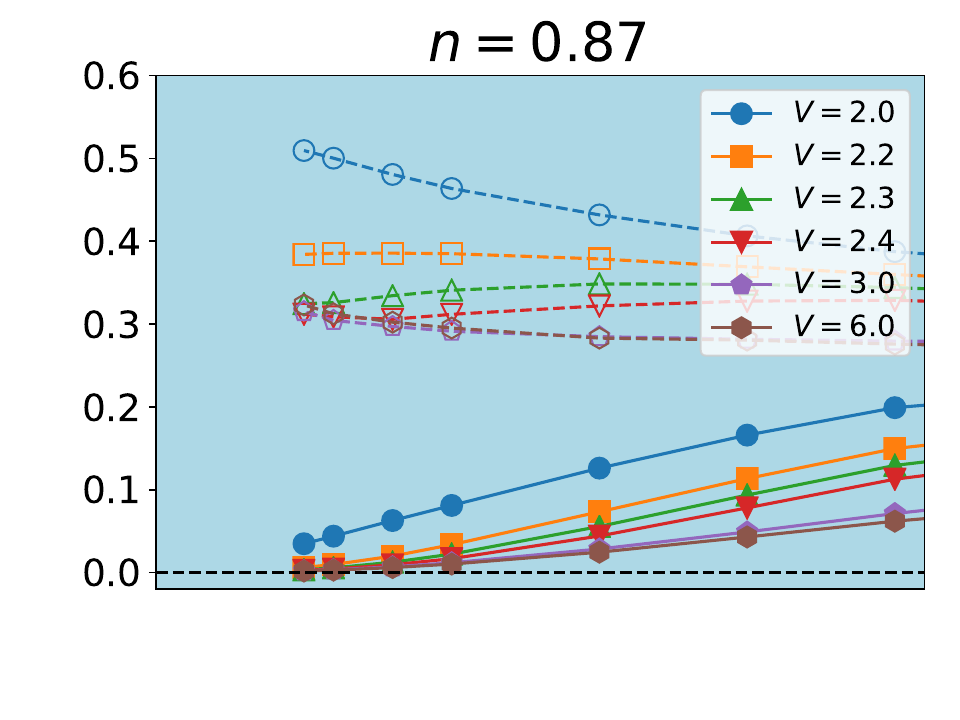,height=4.0cm,width=.32\textwidth, clip}
\hspace{-3mm}
\psfig{figure=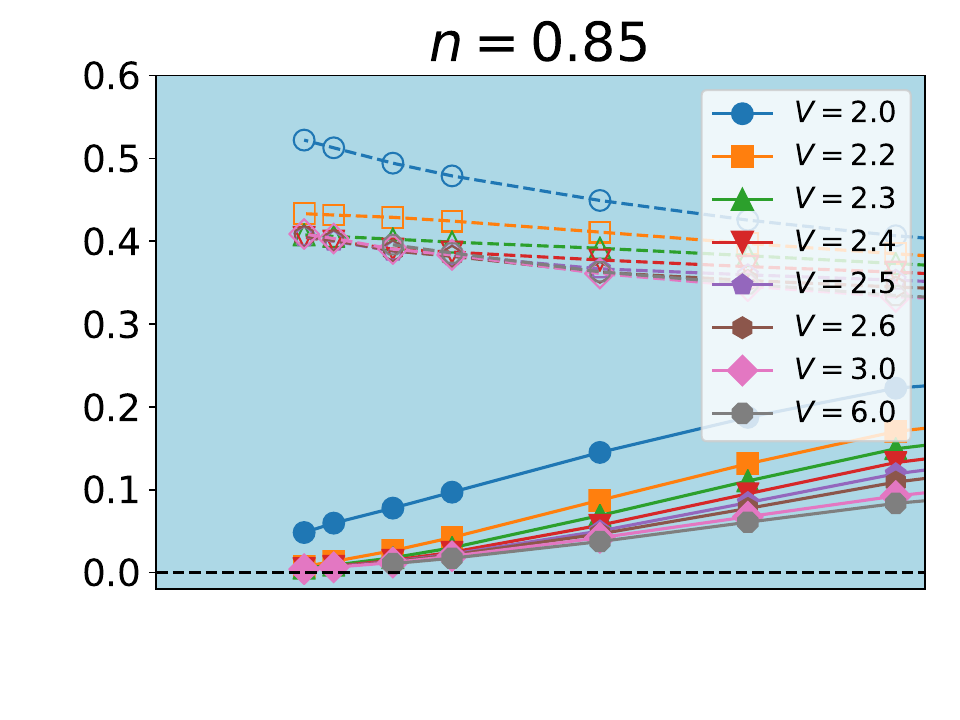,height=4.0cm,width=.32\textwidth, clip}
\psfig{figure=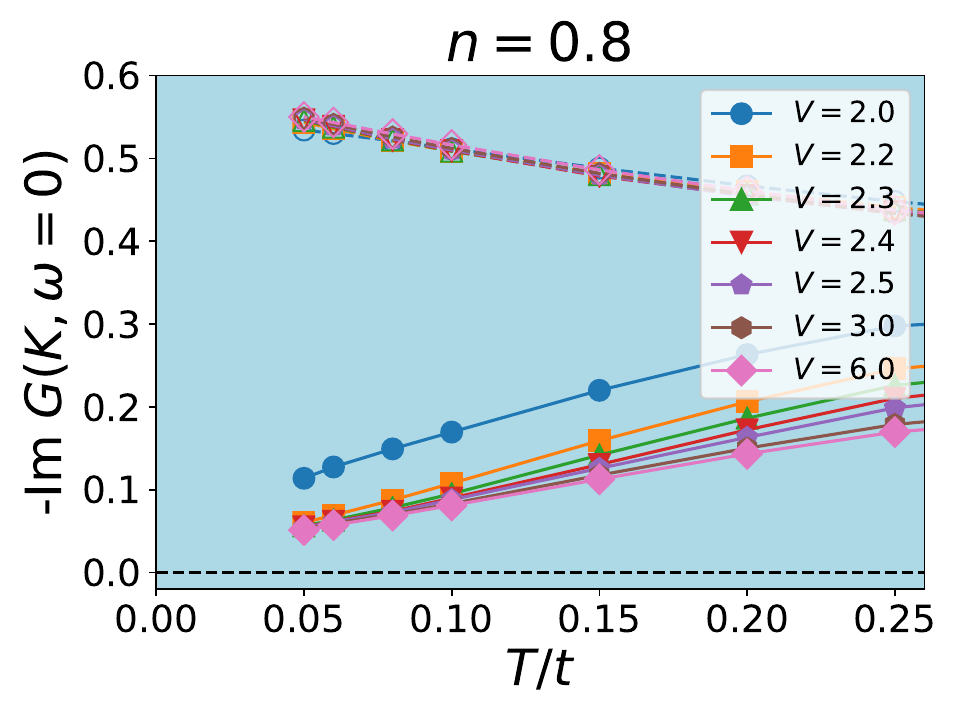,height=4.0cm,width=.32\textwidth, clip}
\hspace{-3mm}
\psfig{figure=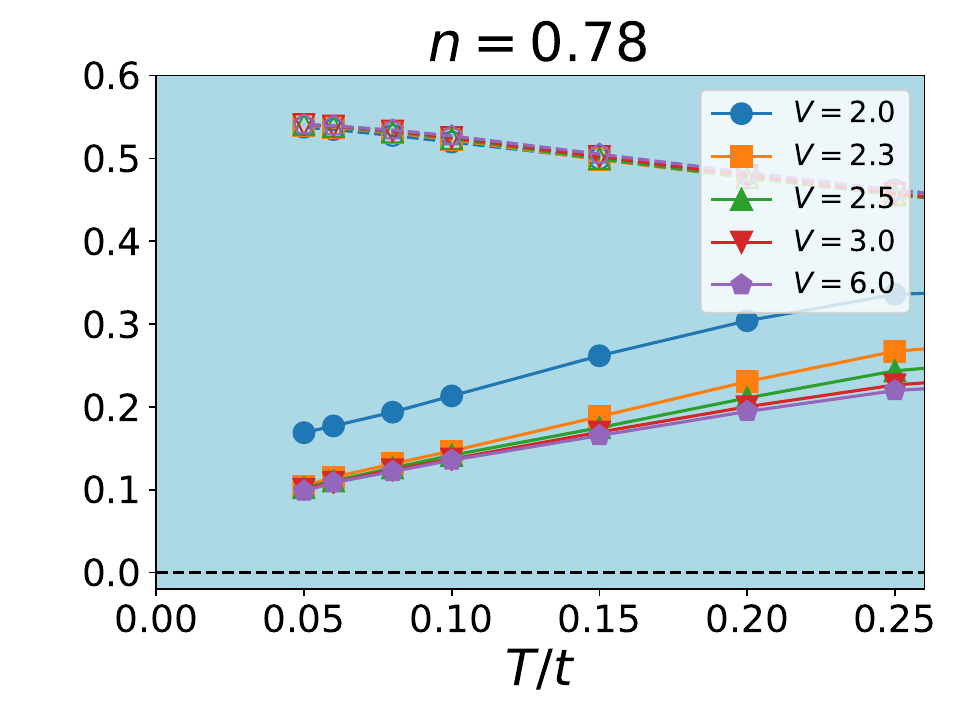,height=4.0cm,width=.32\textwidth, clip}
\hspace{-3mm}
\psfig{figure=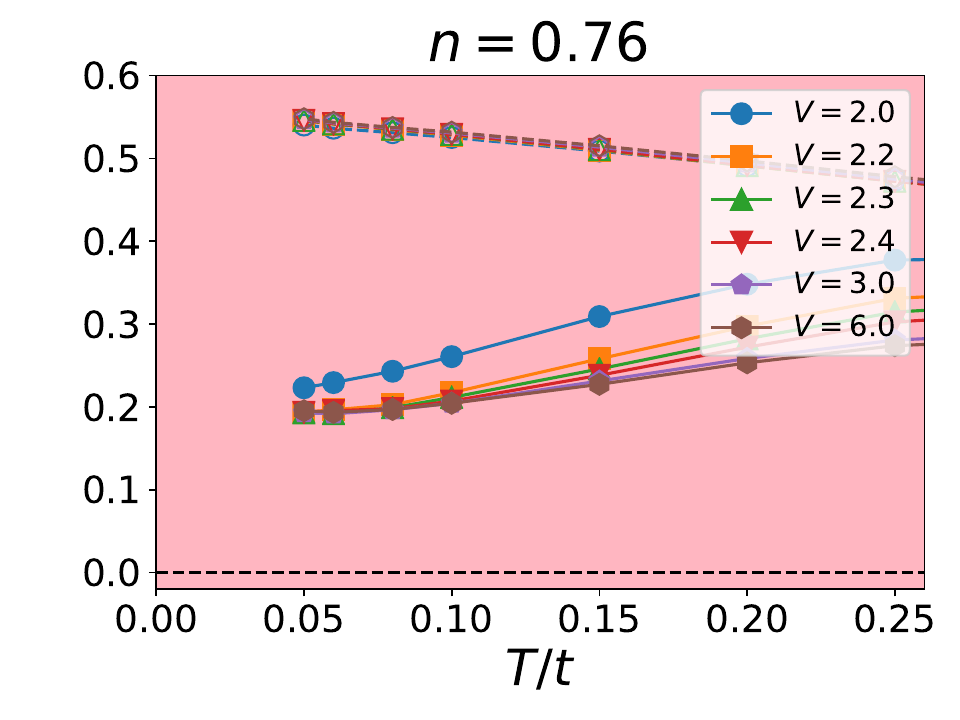,height=4.0cm,width=.32\textwidth, clip}
\caption{Extrapolated zero-frequency -ImG$(\bf{K}, \omega=0)$ obtained from a linear extrapolation of the first two Matsubara frequencies at in-plane antinodal $(\pi,0)$ (solid) and nodal $(\pi/2,\pi/2)$ (dashed) for $\bf K_z=0$.}
\label{ImG}
\end{figure*}

The anisotropy also occurs in the $z$ direction, namely between $\bf K_z=0, \pi$ bands. It can stem from a combination of orbital character and Fermi surface properties~\cite{bilayer5}. In the present bilayer model, there is no orbital effects because of the equivalence between two layers. Hence, the anisotropy should be dominantly driven by the Fermi surface difference associated with $\bf K_z=0, \pi$ bands. As the interlayer hybridization or the doping is increased, the asymmetry between these two bands in the electronic structure and thereby the eigenfunction $\phi_{s^{\pm}}$ gets pronounced. 
Specifically, the frequency dependence is almost compensated between the two bands and approaches zero in the large frequency limit. Previous study~\cite{bilayer5} uncovered that the presence of an incipient band can break this compensation so that the eigenfunctions can approach to a finite value as $\omega \rightarrow \infty$ so that there exists sign change at a finite frequency scale. Here we only observe some vague signals at large dopings and/or larger $V$.

Physically, with the hole doping and/or increasing $V$, the lower energy $\bf K_z=0$ band gradually moves downwards such that the ${s^{\pm}}$ pairing interaction has to change accordingly.
The sign change from negative $\phi_{s^{\pm}}$ at low frequencies to positive at high frequencies is analogous to conventional superconductors with a large Coulomb pseudopotential~\cite{PhysRevLett.100.237001} and the behavior in the extended Hubbard model~\cite{Mi2018}. 
In fact, the $s^{\pm}$-wave nature of the gap reflects an effective pairing interaction for the electron band is attractive at low frequencies, which arises from spin fluctuations instead of conventional electron-phonon coupling. The effective repulsive nature at high energies then arises from the Coulomb interaction.
Correspondingly, the sign change of the eigenfunction manifests the attractive to repulsive change of the effective pairing interaction.

%%%%%%%%%%%%%%%%%%%%%%%%

\subsection{Phase transition at large hybridization $V$ }

The phase diagram Fig.~\ref{phase} depicts the fate of the $s^{\pm}$-wave SC  at strong enough interlayer hybridization $V$ in different doping regime.
From now on, we focus on these transitions by examining the evolution of spectral property via the imaginary part of the Green function. 
To this aim, Fig.~\ref{ImG} summarizes the extrapolated nodal and antinodal zero-frequency -ImG$(\bf{K}, \omega=0)$ as the estimator of the spectral properties, which is obtained from a linear extrapolation of the first two Matsubara frequencies, at in-plane antinodal $\bf{K}=(\pi,0,0)$ (solid) and nodal $(\pi/2,\pi/2,0)$ (dashed) with fixed $\bf{K}_z=0$ for varying densities, where the nodal and antinodal curves show distinctly fruitful features.

\begin{figure}
\psfig{figure=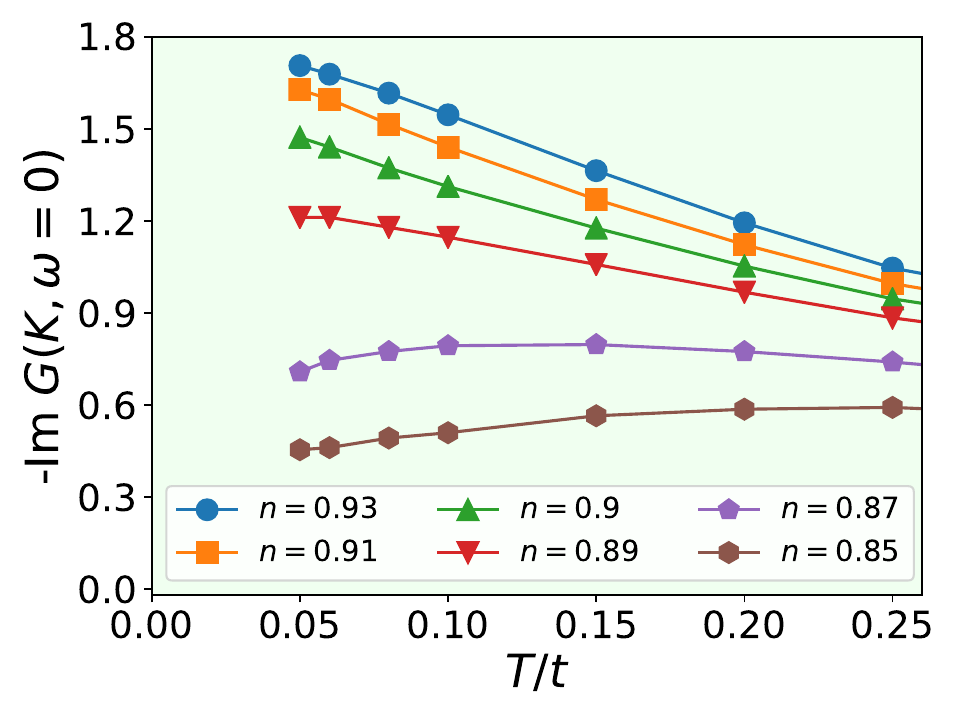,height=5.8cm,width=.47\textwidth, clip}
\caption{Dependence of the extrapolated zero-frequency -ImG$(\bf{K}, \omega=0)$ at in-plane $(\bf{K}_x,\bf{K}_y)=(\pi,\pi)$ with the density for $\bf K_z=0$ at $V/t=3.0$.}
\label{pipi}
\end{figure}

At the underdoped regime e.g. $d=0.93$, increasing the hybridization $V$ results in the gradual downturn of the nodal curves with lowering $T$, which signifies the transition from metallic at smaller $V$ to insulating behavior at sufficiently large $V$. Hence, together with the antinodal curves, this strongly indicates the transition from the pseudogap (PG) to seemingly insulating state. The transition from the highly momentum selective to isotropic behavior is consistent with the $s^{\pm}$-wave eigenfunction shown in Fig.~\ref{eigvec}.
Nonetheless, Fig.~\ref{pipi} indicates that the in-plane $(\bf{K}_x,\bf{K}_y)=(\pi,\pi)$ manifests strong metallic behavior at low dopings, which is weakened by further dopings to show insulating features.
Therefore, this underdoped regime in the phase diagram should be labeled as a correlated metal without the conventional PG features from the deviation between nodal and antinodal directions. This correlated metallicity hosts a strong momentum selectivity arising mainly from other in-plane $\bf{K}$ points like $(\bf{K}_x,\bf{K}_y)=(\pi,\pi)$. In other words, if we extend our definition of PG regime outside of the conventional nodal and antinodal dichotomy, the underdoped regime of the bilayer model can also be viewed as the PG. Here, however, we follow the conventional wording to label it as a separate correlated metal regime in Fig.~\ref{phase}.

Fig.~\ref{ImG} indicates that, with further hole doping, e.g. $d=0.9$, the purely insulating behavior of the nodal direction starts to disappear. For instance, the nodal -ImG$(\bf{K}, \omega=0)$ hesitates to further decrease to lowering $T$ and the $V/t=4.0, 6.0$ curves almost overlap with each other indicating the saturation. 
This change can be more clearly seen at $d=0.89$, where the nodal direction's metallic nature completely deviates from the antinodal's insulating feature, namely the PG physics starts to dominate.

This PG regime extends to around $d=0.78$, beyond which even the nodal direction shows the metallicity so that the whole system transits from PG to metallic. Interestingly, in this overdoped regime, the spectral weight -ImG$(\bf{K}, \omega=0)$ is largely independent on the hybridization over a wide range.
This independence on $V$ results in the vertical phase boundary in Fig.~\ref{phase}, which further implies that the critical densities for the transition between different phases are independent on the hybridization $V$. In other words, the local physics has been fully captured by the moderate hybridization.

%%%%%%%%%%%%%%%%%%%%%%%%

\begin{figure*}[t]
\psfig{figure=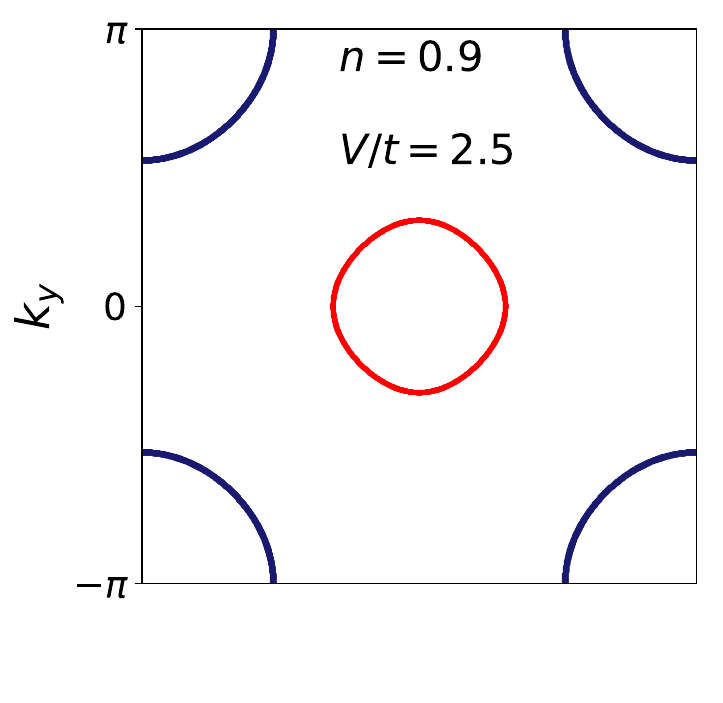,height=4.0cm,width=.24\textwidth, clip}
\vspace{-1em}
\hspace{-0.9em}
\psfig{figure=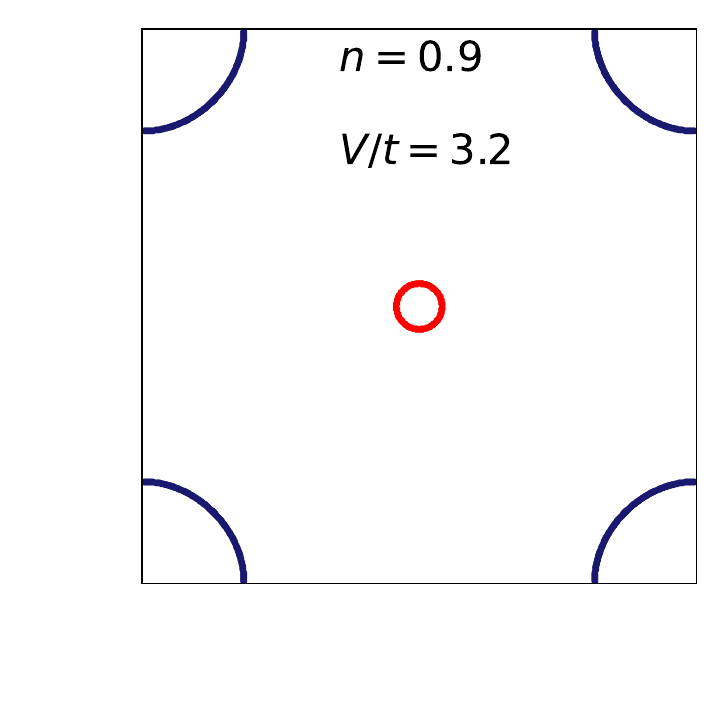,height=4.0cm,width=.24\textwidth, clip}
\hspace{-0.9em}
\psfig{figure=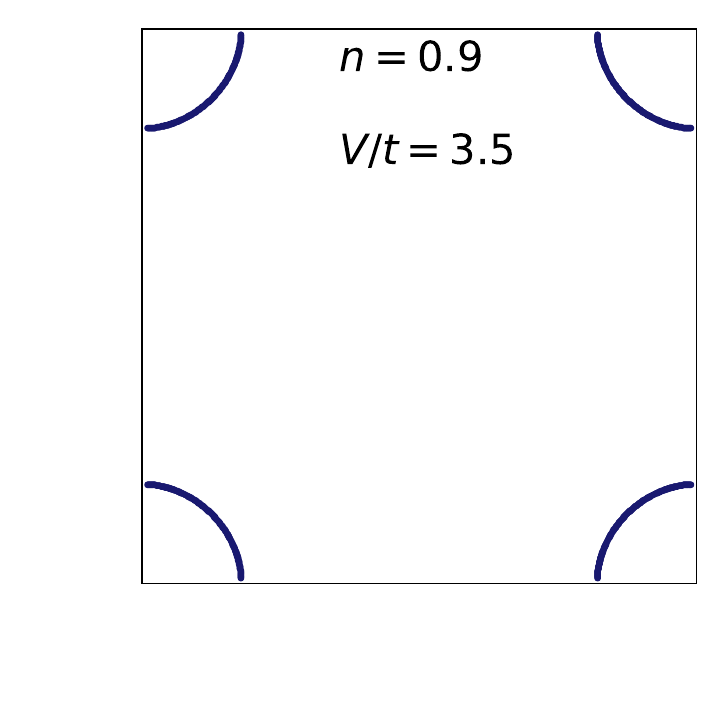,height=4.0cm,width=.24\textwidth, clip}
\hspace{-0.9em}
\psfig{figure=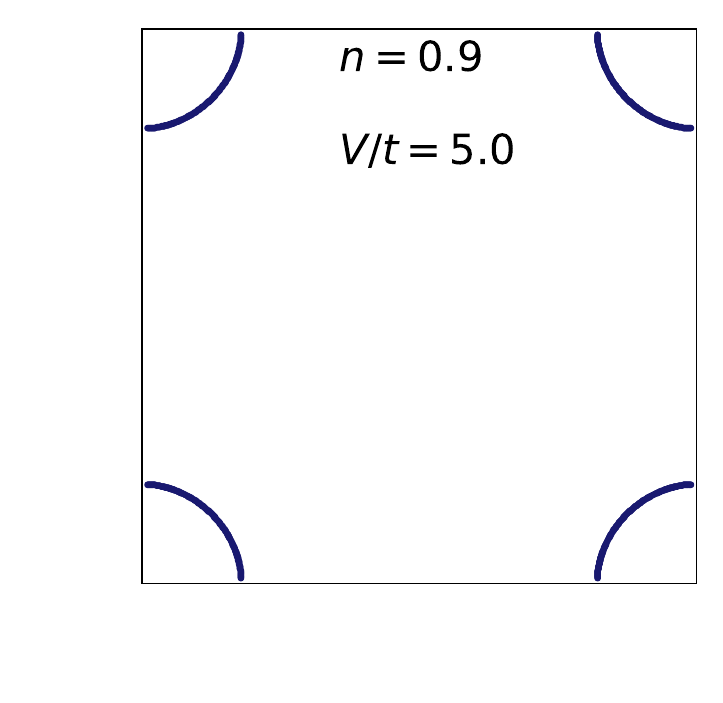,height=4.0cm,width=.24\textwidth, clip}
\psfig{figure=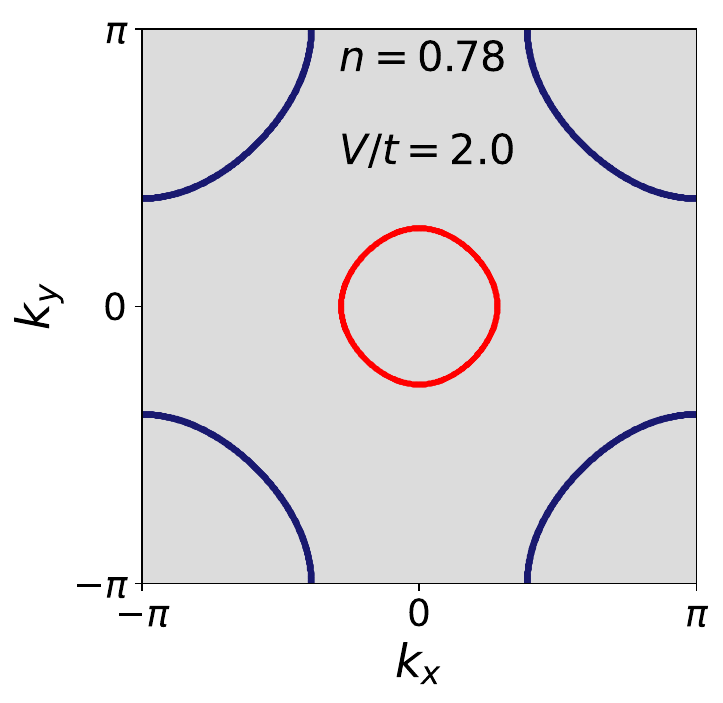,height=4.0cm,width=.24\textwidth, clip}
\hspace{-0.9em}
\psfig{figure=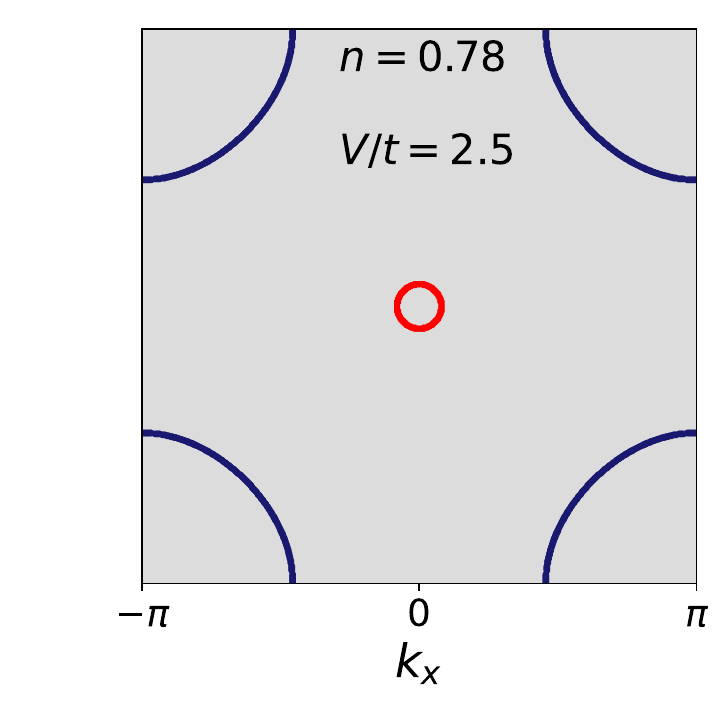,height=4.0cm,width=.24\textwidth, clip}
\hspace{-0.9em}
\psfig{figure=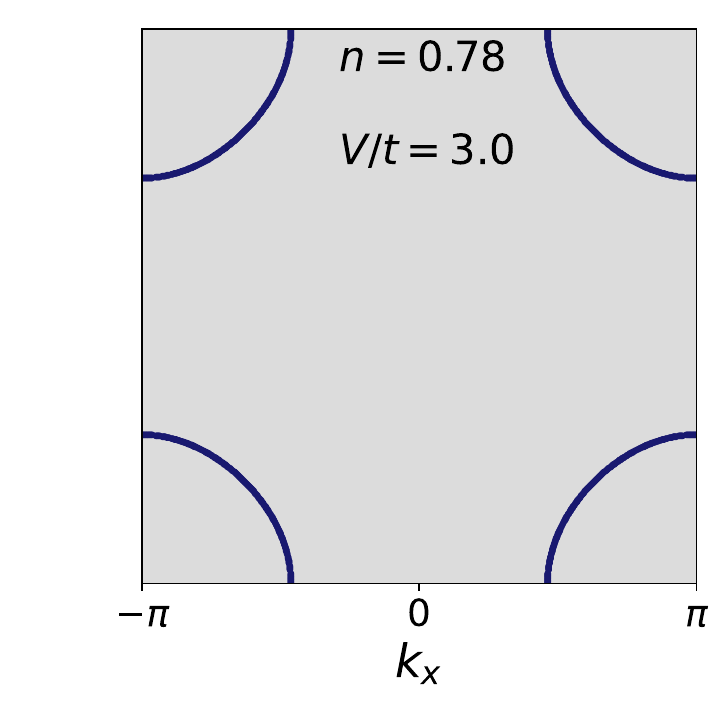,height=4.0cm,width=.24\textwidth, clip}
\hspace{-0.9em}
\psfig{figure=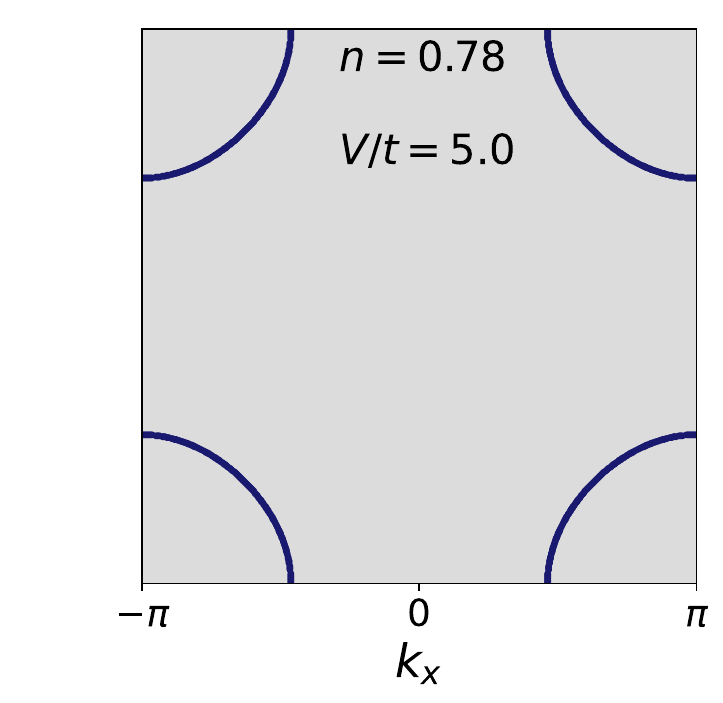,height=4.0cm,width=.24\textwidth, clip}
\caption{Non-interacting Fermi surface for different hybridization $V$ at fixed characteristic (white) $d=0.9$ and (gray) $d=0.78$. The critical $V$ for the disappearance of electron pocket (red circle) for $d=0.78$ is smaller than that for $d=0.9$ matching with the phase diagram Fig.~\ref{phase}.}
\label{FSV}
\end{figure*}

\subsection{Non-interacting Fermi surface}

The phenomena observed in the last sections can be partly understood in the non-interacting limit. 
To this aim, we explore the evolution of the non-interacting Fermi surface (FS) at difference densities and hybridization $V$.

Fig.~\ref{FSV} displays the Fermi surfaces at some typical $V$ at two fixed characteristic density $d=0.9, 0.78$.
The clear shrinking of $k_z=\pi$ FS sheet with increasing $V$ is essential to destroy the inter-band $s^{\pm}$ pairing so that sufficiently large hybridization $V$ always suppresses the SC, as proven in our many-body calculated phase diagram Fig.~\ref{phase}. Note that other densities have qualitatively the same evolution of the FS so that the mechanism of the SC disappearance is closely related to the FS topology change from two sheets to only one that is unfavorable to $s^{\pm}$ pairing.
Another feature lies in that the critical $V$ for the disappearance of electron pocket (red circle) for $d=0.78$ is smaller than that for $d=0.9$, which is consistent with the phase diagram Fig.~\ref{phase}.
In addition, it can be seen from the right panels that the large $V$ does not change FS any more, which can explain the independence of the phase boundary on $V$ in Fig.~\ref{phase} after destroying the SC.

\begin{figure}
\psfig{figure=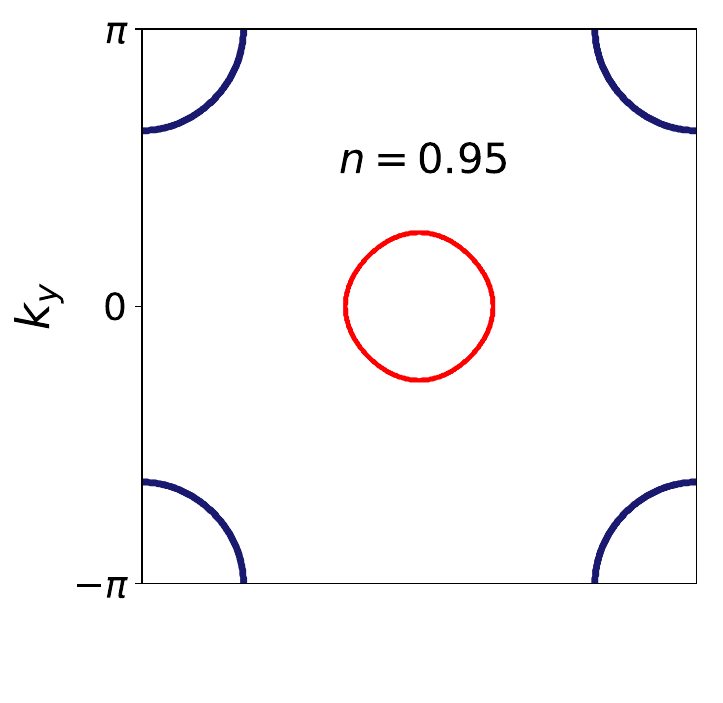,height=3.8cm,width=.22\textwidth, clip}
\vspace{-1em}
\hspace{-0.9em}
\psfig{figure=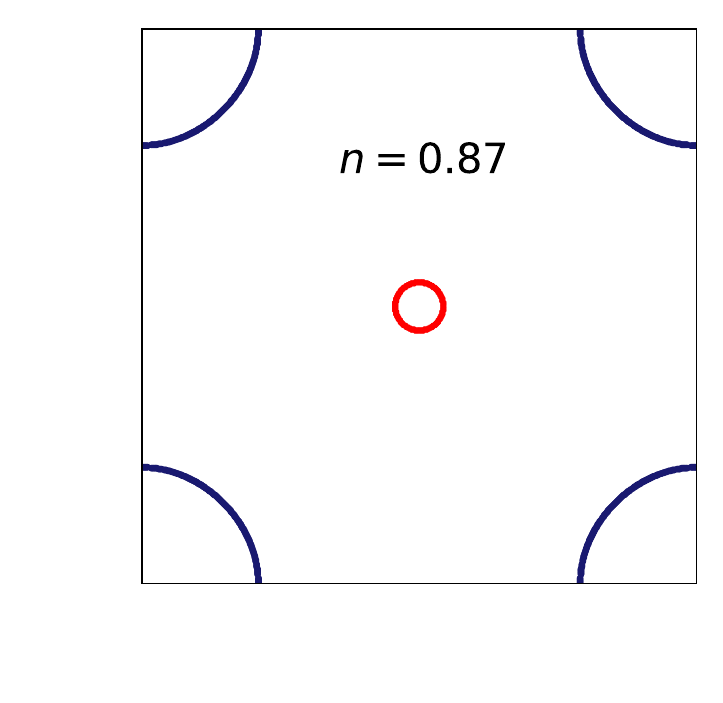,height=3.8cm,width=.22\textwidth, clip}
\psfig{figure=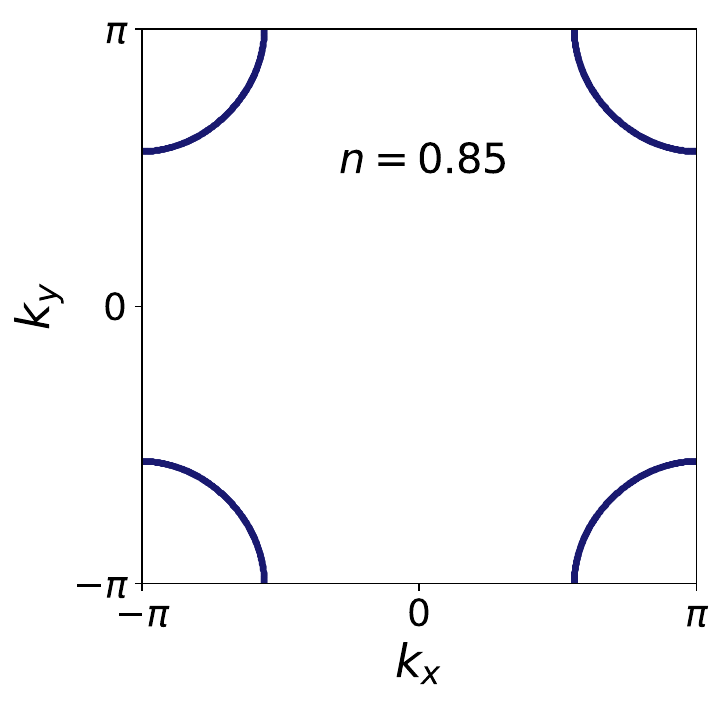,height=3.8cm,width=.22\textwidth, clip}
\hspace{-0.9em}
\psfig{figure=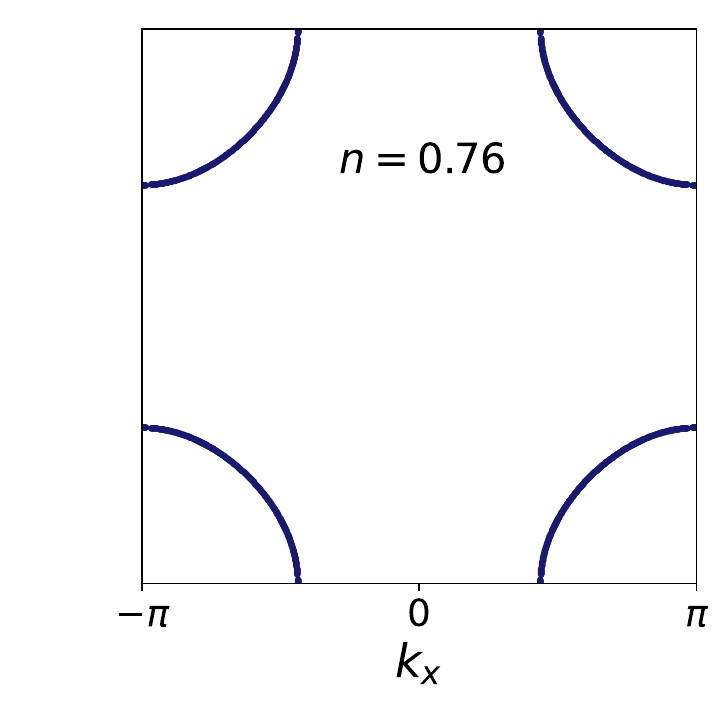,height=3.8cm,width=.22\textwidth, clip}
\caption{Non-interacting Fermi surface for different densities at fixed characteristic hybridization $V/t=3.0$.}
\label{FSd}
\end{figure}

Instead, Fig.~\ref{FSd} illustrates the FS at some typical densities at a fixed characteristic hybridization $V/t=3.0$.
The first two panels still have two FS sheets, which seemingly imply the existence of $s^{\pm}$ SC but this contradicts with our phase diagram Fig.~\ref{phase}. This mismatch can be traced back to be the many-body interaction that can largely reshape the FS.

Generally, the shrinking of FS from two to one sheet is reminiscent of the evolution versus $V$ in Fig.~\ref{FSV}. The difference is reflected in the panels (c-d), where the gradual increasing hole doping is manifested by the enlarging hole pockets and there is no saturation of FS as that of large $V$ limit in Fig.~\ref{FSV}.
Nonetheless, there is no any feature change of FS across the two critical densities around $d=0.91, 0.78$ as shown in Fig.~\ref{phase}. Therefore, it should not be surprising that the two transitions in our phase diagram are not simply FS effects and its full understanding has to resort to the many-body interaction.
We mention that at larger fixed hybridization (not shown), e.g. $V/t=4.0$, there is no electron pocket around $\Gamma$ point even at relatively large density and otherwise the FS evolution with the density is quite similar to $V/t=3.0$ in Fig.~\ref{FSd}.

%%%%%%%%%%%%%%%%%%%%%%%%%%%
\section{Summary and outlook}

Motivated by the recent high pressure experiments on the quantum phase transition from a superconducting state to an insulating-like state in cuprate superconductors as well as the discovered high $T_c \sim 80$ K in rare-earth nickelate La$_3$Ni$_2$O$_7$ with presumably bilayer structure,
we systematically explored the bilayer Hubbard model in the regime of intermediate to large interlayer hybridization to mimic the experimental pressure effects, especially focusing on the fate of $s^{\pm}$-wave superconductivity in the strong interlayer hybridization regime.

By employing dynamic cluster quantum Monte Carlo calculations, our hybridization-density phase diagram Fig.~\ref{phase} indicates that the sufficiently strong interlayer hybridization always destroys the $s^{\pm}$-wave pairing and induces its transition to correlated metallic, pseudogap, or Fermi liquid phases depending on the doping regime. 
The mapped phase diagram is supported by the numerical evidence from the doping and hybridization dependence of the BSE's leading $s^{\pm}$-wave eigenvalue and corresponding eigenvector, the momentum-resolved Green function, and the non-interacting Fermi surface.
In particular, the superconducting $T_c$ of $s^{\pm}$-wave exhibits the dome shape versus the interlayer hybridization.
The uncovered phase diagram Fig.~\ref{phase} is reminiscent of the single-orbital Hubbard model. The major difference lies in the replacement of $d$-wave SC by $s^{\pm}$-wave and vertical phase boundaries. Besides, the interlayer hybridization $V$ plays the role of the temperature scale.

Additoinally, it is found that some physics can already be partly understood via the non-interacting Fermi surface features, particularly the shrinking of electron pocket so that FS topology transits from two to only one sheet, which is unfavorable to $s^{\pm}$ pairing.
However, there is no feature change of FS across the two critical densities around $d=0.91, 0.78$ in our phase diagram. Hence, the two transitions from correlated metal to PG and finally to conventional metallic phase have to originate from the many-body interaction. The FS's evolution at large $V$ versus the doping in Fig.~\ref{FSd} resembles that of the single-band Hubbard model. It is thus reasonable to expect that the physics would then be the same as the single band case and thereby those puzzling transitions between insulating, PG, and metallic behavior of the single-band model persists in the bilayer model at large $V$. Physically, the two nearest sites within two layers combine into a dimer by strong hybridization and consequently serve as a site of the effective single-``orbital'' system on the square lattice.
Therefore, it is plausible to articulate that the bilayer model simply reduces to the single-band model by the strong interlayer hybridization so that the study of bilayer model will benefit the study of the original single-orbital model. 

The rich phase diagram hosted by the bilayer model implies its role as the versatile platform to explore the 2D to 3D crossover physics of Hubbard-type models.
Given that the recent ultracold atomic systems has realized the bilayer Hubbard model~\cite{bilayer4}, the better understanding and especially the direct comparison between the theoretical and experimental observations is highly anticipated. 
Our presented work thus provides valuable insight on the significant role of the interplay of hybridization and electronic interaction in the extensively studied Hubbard models, especially on their potential relevance to the celebrated cuprate and nickelate SC, for instance, if extending to multi-orbitals per layer~\cite{yaoyuan}.

\section{Acknowledgements} 
We acknowledge insightful discussion with Guangming Zhang, Fan Yang, Fa Wang, and Thomas A. Maier.
This work was supported by National Natural Science Foundation of China (NSFC) Grant No. 12174278 and Priority Academic Program Development (PAPD) of Jiangsu Higher Education Institutions.

%%%%%%%%%%%%%%%%%%%%%%%%%%%%%%%%%%%%%%%%%%%%%%%%%%%%%%%%
%\bibliographystyle{plain}
\bibliography{main.bib}

\begin{thebibliography}{32}%
\makeatletter
\providecommand \@ifxundefined [1]{%
 \@ifx{#1\undefined}
}%
\providecommand \@ifnum [1]{%
 \ifnum #1\expandafter \@firstoftwo
 \else \expandafter \@secondoftwo
 \fi
}%
\providecommand \@ifx [1]{%
 \ifx #1\expandafter \@firstoftwo
 \else \expandafter \@secondoftwo
 \fi
}%
\providecommand \natexlab [1]{#1}%
\providecommand \enquote  [1]{``#1''}%
\providecommand \bibnamefont  [1]{#1}%
\providecommand \bibfnamefont [1]{#1}%
\providecommand \citenamefont [1]{#1}%
\providecommand \href@noop [0]{\@secondoftwo}%
\providecommand \href [0]{\begingroup \@sanitize@url \@href}%
\providecommand \@href[1]{\@@startlink{#1}\@@href}%
\providecommand \@@href[1]{\endgroup#1\@@endlink}%
\providecommand \@sanitize@url [0]{\catcode `\\12\catcode `\$12\catcode `\&12\catcode `\#12\catcode `\^12\catcode `\_12\catcode `\%12\relax}%
\providecommand \@@startlink[1]{}%
\providecommand \@@endlink[0]{}%
\providecommand \url  [0]{\begingroup\@sanitize@url \@url }%
\providecommand \@url [1]{\endgroup\@href {#1}{\urlprefix }}%
\providecommand \urlprefix  [0]{URL }%
\providecommand \Eprint [0]{\href }%
\providecommand \doibase [0]{https://doi.org/}%
\providecommand \selectlanguage [0]{\@gobble}%
\providecommand \bibinfo  [0]{\@secondoftwo}%
\providecommand \bibfield  [0]{\@secondoftwo}%
\providecommand \translation [1]{[#1]}%
\providecommand \BibitemOpen [0]{}%
\providecommand \bibitemStop [0]{}%
\providecommand \bibitemNoStop [0]{.\EOS\space}%
\providecommand \EOS [0]{\spacefactor3000\relax}%
\providecommand \BibitemShut  [1]{\csname bibitem#1\endcsname}%
\let\auto@bib@innerbib\@empty
%</preamble>
\bibitem [{\citenamefont {Zhou}\ \emph {et~al.}(2022)\citenamefont {Zhou}, \citenamefont {Guo}, \citenamefont {Cai}, \citenamefont {Zhao}, \citenamefont {Gu}, \citenamefont {Lin}, \citenamefont {Yan}, \citenamefont {Huang}, \citenamefont {Yang}, \citenamefont {Long}, \citenamefont {Gong}, \citenamefont {Li}, \citenamefont {Li}, \citenamefont {Wu}, \citenamefont {Hu}, \citenamefont {Zhou}, \citenamefont {Xiang},\ and\ \citenamefont {Sun}}]{Sunliling2022}%
  \BibitemOpen
  \bibfield  {author} {\bibinfo {author} {\bibfnamefont {Y.}~\bibnamefont {Zhou}}, \bibinfo {author} {\bibfnamefont {J.}~\bibnamefont {Guo}}, \bibinfo {author} {\bibfnamefont {S.}~\bibnamefont {Cai}}, \bibinfo {author} {\bibfnamefont {J.}~\bibnamefont {Zhao}}, \bibinfo {author} {\bibfnamefont {G.}~\bibnamefont {Gu}}, \bibinfo {author} {\bibfnamefont {C.~T.}\ \bibnamefont {Lin}}, \bibinfo {author} {\bibfnamefont {H.}~\bibnamefont {Yan}}, \bibinfo {author} {\bibfnamefont {C.}~\bibnamefont {Huang}}, \bibinfo {author} {\bibfnamefont {C.}~\bibnamefont {Yang}}, \bibinfo {author} {\bibfnamefont {S.}~\bibnamefont {Long}}, \bibinfo {author} {\bibfnamefont {Y.}~\bibnamefont {Gong}}, \bibinfo {author} {\bibfnamefont {Y.}~\bibnamefont {Li}}, \bibinfo {author} {\bibfnamefont {X.}~\bibnamefont {Li}}, \bibinfo {author} {\bibfnamefont {Q.}~\bibnamefont {Wu}}, \bibinfo {author} {\bibfnamefont {J.}~\bibnamefont {Hu}}, \bibinfo {author} {\bibfnamefont {X.}~\bibnamefont {Zhou}}, \bibinfo {author} {\bibfnamefont {T.}~\bibnamefont
  {Xiang}},\ and\ \bibinfo {author} {\bibfnamefont {L.}~\bibnamefont {Sun}},\ }\bibfield  {title} {\bibinfo {title} {Quantum phase transition from superconducting to insulating-like state in a pressurized cuprate superconductor},\ }\href {https://doi.org/10.1038/s41567-022-01513-2} {\bibfield  {journal} {\bibinfo  {journal} {Nature Physics}\ }\textbf {\bibinfo {volume} {18}},\ \bibinfo {pages} {406 } (\bibinfo {year} {2022})}\BibitemShut {NoStop}%
\bibitem [{\citenamefont {Keimer}\ \emph {et~al.}(2015)\citenamefont {Keimer}, \citenamefont {Kivelson}, \citenamefont {Norman}, \citenamefont {Uchida},\ and\ \citenamefont {Zaanen}}]{Keimer2015}%
  \BibitemOpen
  \bibfield  {author} {\bibinfo {author} {\bibfnamefont {B.}~\bibnamefont {Keimer}}, \bibinfo {author} {\bibfnamefont {S.~A.}\ \bibnamefont {Kivelson}}, \bibinfo {author} {\bibfnamefont {M.~R.}\ \bibnamefont {Norman}}, \bibinfo {author} {\bibfnamefont {S.}~\bibnamefont {Uchida}},\ and\ \bibinfo {author} {\bibfnamefont {J.}~\bibnamefont {Zaanen}},\ }\bibfield  {title} {\bibinfo {title} {From quantum matter to high-temperature superconductivity in copper oxides},\ }\href {https://doi.org/10.1038/nature14165} {\bibfield  {journal} {\bibinfo  {journal} {Nature}\ }\textbf {\bibinfo {volume} {518}},\ \bibinfo {pages} {179} (\bibinfo {year} {2015})}\BibitemShut {NoStop}%
\bibitem [{\citenamefont {Bulut}\ \emph {et~al.}(1992)\citenamefont {Bulut}, \citenamefont {Scalapino},\ and\ \citenamefont {Scalettar}}]{bilayer1}%
  \BibitemOpen
  \bibfield  {author} {\bibinfo {author} {\bibfnamefont {N.}~\bibnamefont {Bulut}}, \bibinfo {author} {\bibfnamefont {D.~J.}\ \bibnamefont {Scalapino}},\ and\ \bibinfo {author} {\bibfnamefont {R.~T.}\ \bibnamefont {Scalettar}},\ }\bibfield  {title} {\bibinfo {title} {Nodeless d-wave pairing in a two-layer hubbard model},\ }\href {https://doi.org/10.1103/PhysRevB.45.5577} {\bibfield  {journal} {\bibinfo  {journal} {Phys. Rev. B}\ }\textbf {\bibinfo {volume} {45}},\ \bibinfo {pages} {5577} (\bibinfo {year} {1992})}\BibitemShut {NoStop}%
\bibitem [{\citenamefont {Scalettar}\ \emph {et~al.}(1994)\citenamefont {Scalettar}, \citenamefont {Cannon}, \citenamefont {Scalapino},\ and\ \citenamefont {Sugar}}]{bilayer2}%
  \BibitemOpen
  \bibfield  {author} {\bibinfo {author} {\bibfnamefont {R.~T.}\ \bibnamefont {Scalettar}}, \bibinfo {author} {\bibfnamefont {J.~W.}\ \bibnamefont {Cannon}}, \bibinfo {author} {\bibfnamefont {D.~J.}\ \bibnamefont {Scalapino}},\ and\ \bibinfo {author} {\bibfnamefont {R.~L.}\ \bibnamefont {Sugar}},\ }\bibfield  {title} {\bibinfo {title} {Magnetic and pairing correlations in coupled hubbard planes},\ }\href {https://doi.org/10.1103/PhysRevB.50.13419} {\bibfield  {journal} {\bibinfo  {journal} {Phys. Rev. B}\ }\textbf {\bibinfo {volume} {50}},\ \bibinfo {pages} {13419} (\bibinfo {year} {1994})}\BibitemShut {NoStop}%
\bibitem [{\citenamefont {dos Santos}(1995)}]{bilayer3}%
  \BibitemOpen
  \bibfield  {author} {\bibinfo {author} {\bibfnamefont {R.~R.}\ \bibnamefont {dos Santos}},\ }\bibfield  {title} {\bibinfo {title} {Magnetism and pairing in hubbard bilayers},\ }\href {https://doi.org/10.1103/PhysRevB.51.15540} {\bibfield  {journal} {\bibinfo  {journal} {Phys. Rev. B}\ }\textbf {\bibinfo {volume} {51}},\ \bibinfo {pages} {15540} (\bibinfo {year} {1995})}\BibitemShut {NoStop}%
\bibitem [{\citenamefont {Gall}\ \emph {et~al.}(2021)\citenamefont {Gall}, \citenamefont {Wurz}, \citenamefont {Samland}, \citenamefont {Chan},\ and\ \citenamefont {K{\"o}hl}}]{bilayer4}%
  \BibitemOpen
  \bibfield  {author} {\bibinfo {author} {\bibfnamefont {M.}~\bibnamefont {Gall}}, \bibinfo {author} {\bibfnamefont {N.}~\bibnamefont {Wurz}}, \bibinfo {author} {\bibfnamefont {J.}~\bibnamefont {Samland}}, \bibinfo {author} {\bibfnamefont {C.~F.}\ \bibnamefont {Chan}},\ and\ \bibinfo {author} {\bibfnamefont {M.}~\bibnamefont {K{\"o}hl}},\ }\bibfield  {title} {\bibinfo {title} {Competing magnetic orders in a bilayer hubbard model with ultracold atoms},\ }\href {https://doi.org/10.1038/s41586-020-03058-x} {\bibfield  {journal} {\bibinfo  {journal} {Nature}\ }\textbf {\bibinfo {volume} {589}},\ \bibinfo {pages} {40 } (\bibinfo {year} {2021})}\BibitemShut {NoStop}%
\bibitem [{\citenamefont {Karakuzu}\ \emph {et~al.}(2021)\citenamefont {Karakuzu}, \citenamefont {Johnston},\ and\ \citenamefont {Maier}}]{bilayer5}%
  \BibitemOpen
  \bibfield  {author} {\bibinfo {author} {\bibfnamefont {S.}~\bibnamefont {Karakuzu}}, \bibinfo {author} {\bibfnamefont {S.}~\bibnamefont {Johnston}},\ and\ \bibinfo {author} {\bibfnamefont {T.~A.}\ \bibnamefont {Maier}},\ }\bibfield  {title} {\bibinfo {title} {Superconductivity in the bilayer hubbard model: Two fermi surfaces are better than one},\ }\href {https://doi.org/10.1103/PhysRevB.104.245109} {\bibfield  {journal} {\bibinfo  {journal} {Phys. Rev. B}\ }\textbf {\bibinfo {volume} {104}},\ \bibinfo {pages} {245109} (\bibinfo {year} {2021})}\BibitemShut {NoStop}%
\bibitem [{\citenamefont {Vanhala}\ \emph {et~al.}(2015)\citenamefont {Vanhala}, \citenamefont {Baarsma}, \citenamefont {Heikkinen}, \citenamefont {Troyer}, \citenamefont {Harju},\ and\ \citenamefont {T\"orm\"a}}]{bilayer6}%
  \BibitemOpen
  \bibfield  {author} {\bibinfo {author} {\bibfnamefont {T.~I.}\ \bibnamefont {Vanhala}}, \bibinfo {author} {\bibfnamefont {J.~E.}\ \bibnamefont {Baarsma}}, \bibinfo {author} {\bibfnamefont {M.~O.~J.}\ \bibnamefont {Heikkinen}}, \bibinfo {author} {\bibfnamefont {M.}~\bibnamefont {Troyer}}, \bibinfo {author} {\bibfnamefont {A.}~\bibnamefont {Harju}},\ and\ \bibinfo {author} {\bibfnamefont {P.}~\bibnamefont {T\"orm\"a}},\ }\bibfield  {title} {\bibinfo {title} {Superfluidity and density order in a bilayer extended hubbard model},\ }\href {https://doi.org/10.1103/PhysRevB.91.144510} {\bibfield  {journal} {\bibinfo  {journal} {Phys. Rev. B}\ }\textbf {\bibinfo {volume} {91}},\ \bibinfo {pages} {144510} (\bibinfo {year} {2015})}\BibitemShut {NoStop}%
\bibitem [{\citenamefont {Golor}\ and\ \citenamefont {Wessel}(2015)}]{bilayer7}%
  \BibitemOpen
  \bibfield  {author} {\bibinfo {author} {\bibfnamefont {M.}~\bibnamefont {Golor}}\ and\ \bibinfo {author} {\bibfnamefont {S.}~\bibnamefont {Wessel}},\ }\bibfield  {title} {\bibinfo {title} {Nonlocal density interactions in auxiliary-field quantum monte carlo simulations: Application to the square lattice bilayer and honeycomb lattice},\ }\href {https://doi.org/10.1103/PhysRevB.92.195154} {\bibfield  {journal} {\bibinfo  {journal} {Phys. Rev. B}\ }\textbf {\bibinfo {volume} {92}},\ \bibinfo {pages} {195154} (\bibinfo {year} {2015})}\BibitemShut {NoStop}%
\bibitem [{\citenamefont {Maier}\ and\ \citenamefont {Scalapino}(2011)}]{bilayer8}%
  \BibitemOpen
  \bibfield  {author} {\bibinfo {author} {\bibfnamefont {T.~A.}\ \bibnamefont {Maier}}\ and\ \bibinfo {author} {\bibfnamefont {D.~J.}\ \bibnamefont {Scalapino}},\ }\bibfield  {title} {\bibinfo {title} {Pair structure and the pairing interaction in a bilayer hubbard model for unconventional superconductivity},\ }\href {https://doi.org/10.1103/PhysRevB.84.180513} {\bibfield  {journal} {\bibinfo  {journal} {Phys. Rev. B}\ }\textbf {\bibinfo {volume} {84}},\ \bibinfo {pages} {180513} (\bibinfo {year} {2011})}\BibitemShut {NoStop}%
\bibitem [{\citenamefont {Ochi}\ \emph {et~al.}(2022)\citenamefont {Ochi}, \citenamefont {Tajima}, \citenamefont {Iida},\ and\ \citenamefont {Aoki}}]{incipient2}%
  \BibitemOpen
  \bibfield  {author} {\bibinfo {author} {\bibfnamefont {K.}~\bibnamefont {Ochi}}, \bibinfo {author} {\bibfnamefont {H.}~\bibnamefont {Tajima}}, \bibinfo {author} {\bibfnamefont {K.}~\bibnamefont {Iida}},\ and\ \bibinfo {author} {\bibfnamefont {H.}~\bibnamefont {Aoki}},\ }\bibfield  {title} {\bibinfo {title} {Resonant pair-exchange scattering and bcs-bec crossover in a system composed of dispersive and heavy incipient bands: A feshbach analogy},\ }\href {https://doi.org/10.1103/PhysRevResearch.4.013032} {\bibfield  {journal} {\bibinfo  {journal} {Phys. Rev. Res.}\ }\textbf {\bibinfo {volume} {4}},\ \bibinfo {pages} {013032} (\bibinfo {year} {2022})}\BibitemShut {NoStop}%
\bibitem [{\citenamefont {Linscheid}\ \emph {et~al.}(2016)\citenamefont {Linscheid}, \citenamefont {Maiti}, \citenamefont {Wang}, \citenamefont {Johnston},\ and\ \citenamefont {Hirschfeld}}]{incipient3}%
  \BibitemOpen
  \bibfield  {author} {\bibinfo {author} {\bibfnamefont {A.}~\bibnamefont {Linscheid}}, \bibinfo {author} {\bibfnamefont {S.}~\bibnamefont {Maiti}}, \bibinfo {author} {\bibfnamefont {Y.}~\bibnamefont {Wang}}, \bibinfo {author} {\bibfnamefont {S.}~\bibnamefont {Johnston}},\ and\ \bibinfo {author} {\bibfnamefont {P.~J.}\ \bibnamefont {Hirschfeld}},\ }\bibfield  {title} {\bibinfo {title} {High ${T}_{c}$ via spin fluctuations from incipient bands: Application to monolayers and intercalates of fese},\ }\href {https://doi.org/10.1103/PhysRevLett.117.077003} {\bibfield  {journal} {\bibinfo  {journal} {Phys. Rev. Lett.}\ }\textbf {\bibinfo {volume} {117}},\ \bibinfo {pages} {077003} (\bibinfo {year} {2016})}\BibitemShut {NoStop}%
\bibitem [{\citenamefont {Yue}\ \emph {et~al.}(2022)\citenamefont {Yue}, \citenamefont {Aoki},\ and\ \citenamefont {Werner}}]{Werner}%
  \BibitemOpen
  \bibfield  {author} {\bibinfo {author} {\bibfnamefont {C.}~\bibnamefont {Yue}}, \bibinfo {author} {\bibfnamefont {H.}~\bibnamefont {Aoki}},\ and\ \bibinfo {author} {\bibfnamefont {P.}~\bibnamefont {Werner}},\ }\bibfield  {title} {\bibinfo {title} {Superconductivity enhanced by pair fluctuations between wide and narrow bands},\ }\href {https://doi.org/10.1103/PhysRevB.106.L180506} {\bibfield  {journal} {\bibinfo  {journal} {Phys. Rev. B}\ }\textbf {\bibinfo {volume} {106}},\ \bibinfo {pages} {L180506} (\bibinfo {year} {2022})}\BibitemShut {NoStop}%
\bibitem [{\citenamefont {Dee}\ \emph {et~al.}(2022)\citenamefont {Dee}, \citenamefont {Johnston},\ and\ \citenamefont {Maier}}]{Maier2022}%
  \BibitemOpen
  \bibfield  {author} {\bibinfo {author} {\bibfnamefont {P.~M.}\ \bibnamefont {Dee}}, \bibinfo {author} {\bibfnamefont {S.}~\bibnamefont {Johnston}},\ and\ \bibinfo {author} {\bibfnamefont {T.~A.}\ \bibnamefont {Maier}},\ }\bibfield  {title} {\bibinfo {title} {Enhancing ${T}_{\mathrm{c}}$ in a composite superconductor/metal bilayer system: A dynamical cluster approximation study},\ }\href {https://doi.org/10.1103/PhysRevB.105.214502} {\bibfield  {journal} {\bibinfo  {journal} {Phys. Rev. B}\ }\textbf {\bibinfo {volume} {105}},\ \bibinfo {pages} {214502} (\bibinfo {year} {2022})}\BibitemShut {NoStop}%
\bibitem [{\citenamefont {Sun}\ \emph {et~al.}(2023)\citenamefont {Sun}, \citenamefont {Huo}, \citenamefont {Hu}, \citenamefont {Li}, \citenamefont {Liu}, \citenamefont {Han}, \citenamefont {Tang}, \citenamefont {Mao}, \citenamefont {Yang}, \citenamefont {Wang} \emph {et~al.}}]{327}%
  \BibitemOpen
  \bibfield  {author} {\bibinfo {author} {\bibfnamefont {H.}~\bibnamefont {Sun}}, \bibinfo {author} {\bibfnamefont {M.}~\bibnamefont {Huo}}, \bibinfo {author} {\bibfnamefont {X.}~\bibnamefont {Hu}}, \bibinfo {author} {\bibfnamefont {J.}~\bibnamefont {Li}}, \bibinfo {author} {\bibfnamefont {Z.}~\bibnamefont {Liu}}, \bibinfo {author} {\bibfnamefont {Y.}~\bibnamefont {Han}}, \bibinfo {author} {\bibfnamefont {L.}~\bibnamefont {Tang}}, \bibinfo {author} {\bibfnamefont {Z.}~\bibnamefont {Mao}}, \bibinfo {author} {\bibfnamefont {P.}~\bibnamefont {Yang}}, \bibinfo {author} {\bibfnamefont {B.}~\bibnamefont {Wang}}, \emph {et~al.},\ }\bibfield  {title} {\bibinfo {title} {Signatures of superconductivity near 80 k in a nickelate under high pressure},\ }\href {https://doi.org/10.1038/s41586-023-06408-7} {\bibfield  {journal} {\bibinfo  {journal} {Nature}\ }\textbf {\bibinfo {volume} {621}},\ \bibinfo {pages} {493} (\bibinfo {year} {2023})}\BibitemShut {NoStop}%
\bibitem [{\citenamefont {Qin}\ and\ \citenamefont {Jiang}(2023)}]{Mi23}%
  \BibitemOpen
  \bibfield  {author} {\bibinfo {author} {\bibfnamefont {C.}~\bibnamefont {Qin}}\ and\ \bibinfo {author} {\bibfnamefont {M.}~\bibnamefont {Jiang}},\ }\bibfield  {title} {\bibinfo {title} {Inversion symmetry breaking in a bilayer multiorbital hubbard model with impurity approximation},\ }\href {https://doi.org/10.1103/PhysRevB.108.085137} {\bibfield  {journal} {\bibinfo  {journal} {Phys. Rev. B}\ }\textbf {\bibinfo {volume} {108}},\ \bibinfo {pages} {085137} (\bibinfo {year} {2023})}\BibitemShut {NoStop}%
\bibitem [{\citenamefont {Li}\ \emph {et~al.}(2019)\citenamefont {Li}, \citenamefont {Lee}, \citenamefont {Wang}, \citenamefont {Osada}, \citenamefont {Crossley}, \citenamefont {Lee}, \citenamefont {Cui}, \citenamefont {Hikita},\ and\ \citenamefont {Hwang}}]{2019Nature}%
  \BibitemOpen
  \bibfield  {author} {\bibinfo {author} {\bibfnamefont {D.}~\bibnamefont {Li}}, \bibinfo {author} {\bibfnamefont {K.}~\bibnamefont {Lee}}, \bibinfo {author} {\bibfnamefont {B.~Y.}\ \bibnamefont {Wang}}, \bibinfo {author} {\bibfnamefont {M.}~\bibnamefont {Osada}}, \bibinfo {author} {\bibfnamefont {S.}~\bibnamefont {Crossley}}, \bibinfo {author} {\bibfnamefont {H.~R.}\ \bibnamefont {Lee}}, \bibinfo {author} {\bibfnamefont {Y.}~\bibnamefont {Cui}}, \bibinfo {author} {\bibfnamefont {Y.}~\bibnamefont {Hikita}},\ and\ \bibinfo {author} {\bibfnamefont {H.~Y.}\ \bibnamefont {Hwang}},\ }\bibfield  {title} {\bibinfo {title} {Superconductivity in an infinite-layer nickelate},\ }\href {https://doi.org/10.1038/s41586-019-1496-5} {\bibfield  {journal} {\bibinfo  {journal} {Nature}\ }\textbf {\bibinfo {volume} {572}},\ \bibinfo {pages} {624 } (\bibinfo {year} {2019})}\BibitemShut {NoStop}%
\bibitem [{\citenamefont {Nomura}\ and\ \citenamefont {Arita}(2022)}]{Aritareview}%
  \BibitemOpen
  \bibfield  {author} {\bibinfo {author} {\bibfnamefont {Y.}~\bibnamefont {Nomura}}\ and\ \bibinfo {author} {\bibfnamefont {R.}~\bibnamefont {Arita}},\ }\bibfield  {title} {\bibinfo {title} {Superconductivity in infinite-layer nickelates},\ }\href {https://dx.doi.org/10.1088/1361-6633/ac5a60} {\bibfield  {journal} {\bibinfo  {journal} {Reports on Progress in Physics}\ }\textbf {\bibinfo {volume} {85}},\ \bibinfo {pages} {052501} (\bibinfo {year} {2022})}\BibitemShut {NoStop}%
\bibitem [{\citenamefont {Botana}\ \emph {et~al.}(2020)\citenamefont {Botana}, \citenamefont {Bernardini},\ and\ \citenamefont {Cano}}]{Botana_review}%
  \BibitemOpen
  \bibfield  {author} {\bibinfo {author} {\bibfnamefont {A.~S.}\ \bibnamefont {Botana}}, \bibinfo {author} {\bibfnamefont {F.}~\bibnamefont {Bernardini}},\ and\ \bibinfo {author} {\bibfnamefont {A.}~\bibnamefont {Cano}},\ }\bibfield  {title} {\bibinfo {title} {Nickelate superconductors: An ongoing dialog between theory and experiments},\ }\href {https://doi.org/10.1134/S1063776121040026} {\bibfield  {journal} {\bibinfo  {journal} {Journal of Experimental and Theoretical Physics}\ }\textbf {\bibinfo {volume} {132}},\ \bibinfo {pages} {618} (\bibinfo {year} {2020})}\BibitemShut {NoStop}%
\bibitem [{\citenamefont {Held}\ \emph {et~al.}(2022)\citenamefont {Held}, \citenamefont {Si}, \citenamefont {Worm}, \citenamefont {Janson}, \citenamefont {Arita}, \citenamefont {Zhong}, \citenamefont {Tomczak},\ and\ \citenamefont {Kitatani}}]{Held2022}%
  \BibitemOpen
  \bibfield  {author} {\bibinfo {author} {\bibfnamefont {K.}~\bibnamefont {Held}}, \bibinfo {author} {\bibfnamefont {L.}~\bibnamefont {Si}}, \bibinfo {author} {\bibfnamefont {P.}~\bibnamefont {Worm}}, \bibinfo {author} {\bibfnamefont {O.}~\bibnamefont {Janson}}, \bibinfo {author} {\bibfnamefont {R.}~\bibnamefont {Arita}}, \bibinfo {author} {\bibfnamefont {Z.}~\bibnamefont {Zhong}}, \bibinfo {author} {\bibfnamefont {J.~M.}\ \bibnamefont {Tomczak}},\ and\ \bibinfo {author} {\bibfnamefont {M.}~\bibnamefont {Kitatani}},\ }\bibfield  {title} {\bibinfo {title} {Phase diagram of nickelate superconductors calculated by dynamical vertex approximation},\ }\href {https://doi.org/10.3389/fphy.2021.810394} {\bibfield  {journal} {\bibinfo  {journal} {Frontiers in Physics}\ }\textbf {\bibinfo {volume} {9}},\ \bibinfo {pages} {810394} (\bibinfo {year} {2022})}\BibitemShut {NoStop}%
\bibitem [{\citenamefont {Chen}\ \emph {et~al.}(2022)\citenamefont {Chen}, \citenamefont {Hampel}, \citenamefont {Karp}, \citenamefont {Lechermann},\ and\ \citenamefont {Millis}}]{Hanghuireview}%
  \BibitemOpen
  \bibfield  {author} {\bibinfo {author} {\bibfnamefont {H.}~\bibnamefont {Chen}}, \bibinfo {author} {\bibfnamefont {A.}~\bibnamefont {Hampel}}, \bibinfo {author} {\bibfnamefont {J.}~\bibnamefont {Karp}}, \bibinfo {author} {\bibfnamefont {F.}~\bibnamefont {Lechermann}},\ and\ \bibinfo {author} {\bibfnamefont {A.~J.}\ \bibnamefont {Millis}},\ }\bibfield  {title} {\bibinfo {title} {Dynamical mean field studies of infinite layer nickelates: Physics results and methodological implications},\ }\href {https://www.frontiersin.org/journals/physics/articles/10.3389/fphy.2022.835942} {\bibfield  {journal} {\bibinfo  {journal} {Frontiers in Physics}\ }\textbf {\bibinfo {volume} {10}},\ \bibinfo {pages} {835942} (\bibinfo {year} {2022})}\BibitemShut {NoStop}%
\bibitem [{\citenamefont {Hettler}\ \emph {et~al.}(1998)\citenamefont {Hettler}, \citenamefont {Tahvildar-Zadeh}, \citenamefont {Jarrell}, \citenamefont {Pruschke},\ and\ \citenamefont {Krishnamurthy}}]{Hettler98}%
  \BibitemOpen
  \bibfield  {author} {\bibinfo {author} {\bibfnamefont {M.~H.}\ \bibnamefont {Hettler}}, \bibinfo {author} {\bibfnamefont {A.~N.}\ \bibnamefont {Tahvildar-Zadeh}}, \bibinfo {author} {\bibfnamefont {M.}~\bibnamefont {Jarrell}}, \bibinfo {author} {\bibfnamefont {T.}~\bibnamefont {Pruschke}},\ and\ \bibinfo {author} {\bibfnamefont {H.~R.}\ \bibnamefont {Krishnamurthy}},\ }\bibfield  {title} {\bibinfo {title} {Nonlocal dynamical correlations of strongly interacting electron systems},\ }\href {https://doi.org/10.1103/PhysRevB.58.R7475} {\bibfield  {journal} {\bibinfo  {journal} {Phys. Rev. B}\ }\textbf {\bibinfo {volume} {58}},\ \bibinfo {pages} {R7475} (\bibinfo {year} {1998})}\BibitemShut {NoStop}%
\bibitem [{\citenamefont {Maier}\ \emph {et~al.}(2005)\citenamefont {Maier}, \citenamefont {Jarrell}, \citenamefont {Pruschke},\ and\ \citenamefont {Hettler}}]{Maier05}%
  \BibitemOpen
  \bibfield  {author} {\bibinfo {author} {\bibfnamefont {T.}~\bibnamefont {Maier}}, \bibinfo {author} {\bibfnamefont {M.}~\bibnamefont {Jarrell}}, \bibinfo {author} {\bibfnamefont {T.}~\bibnamefont {Pruschke}},\ and\ \bibinfo {author} {\bibfnamefont {M.~H.}\ \bibnamefont {Hettler}},\ }\bibfield  {title} {\bibinfo {title} {Quantum cluster theories},\ }\href {https://doi.org/10.1103/RevModPhys.77.1027} {\bibfield  {journal} {\bibinfo  {journal} {Rev. Mod. Phys.}\ }\textbf {\bibinfo {volume} {77}},\ \bibinfo {pages} {1027} (\bibinfo {year} {2005})}\BibitemShut {NoStop}%
\bibitem [{\citenamefont {Hähner}\ \emph {et~al.}(2020)\citenamefont {Hähner}, \citenamefont {Alvarez}, \citenamefont {Maier}, \citenamefont {Solcà}, \citenamefont {Staar}, \citenamefont {Summers},\ and\ \citenamefont {Schulthess}}]{code}%
  \BibitemOpen
  \bibfield  {author} {\bibinfo {author} {\bibfnamefont {U.~R.}\ \bibnamefont {Hähner}}, \bibinfo {author} {\bibfnamefont {G.}~\bibnamefont {Alvarez}}, \bibinfo {author} {\bibfnamefont {T.~A.}\ \bibnamefont {Maier}}, \bibinfo {author} {\bibfnamefont {R.}~\bibnamefont {Solcà}}, \bibinfo {author} {\bibfnamefont {P.}~\bibnamefont {Staar}}, \bibinfo {author} {\bibfnamefont {M.~S.}\ \bibnamefont {Summers}},\ and\ \bibinfo {author} {\bibfnamefont {T.~C.}\ \bibnamefont {Schulthess}},\ }\bibfield  {title} {\bibinfo {title} {Dca++: A software framework to solve correlated electron problems with modern quantum cluster methods},\ }\href {https://doi.org/https://doi.org/10.1016/j.cpc.2019.01.006} {\bibfield  {journal} {\bibinfo  {journal} {Computer Physics Communications}\ }\textbf {\bibinfo {volume} {246}},\ \bibinfo {pages} {106709} (\bibinfo {year} {2020})}\BibitemShut {NoStop}%
\bibitem [{\citenamefont {Gull}\ \emph {et~al.}(2008)\citenamefont {Gull}, \citenamefont {Werner}, \citenamefont {Parcollet},\ and\ \citenamefont {Troyer}}]{GullCTAUX}%
  \BibitemOpen
  \bibfield  {author} {\bibinfo {author} {\bibfnamefont {E.}~\bibnamefont {Gull}}, \bibinfo {author} {\bibfnamefont {P.}~\bibnamefont {Werner}}, \bibinfo {author} {\bibfnamefont {O.}~\bibnamefont {Parcollet}},\ and\ \bibinfo {author} {\bibfnamefont {M.}~\bibnamefont {Troyer}},\ }\bibfield  {title} {\bibinfo {title} {Continuous-time auxiliary-field monte carlo for quantum impurity models},\ }\href {https://doi.org/10.1209/0295-5075/82/57003} {\bibfield  {journal} {\bibinfo  {journal} {Europhysics Letters}\ }\textbf {\bibinfo {volume} {82}},\ \bibinfo {pages} {57003} (\bibinfo {year} {2008})}\BibitemShut {NoStop}%
\bibitem [{\citenamefont {Maier}\ \emph {et~al.}(2006)\citenamefont {Maier}, \citenamefont {Jarrell},\ and\ \citenamefont {Scalapino}}]{Maier2006}%
  \BibitemOpen
  \bibfield  {author} {\bibinfo {author} {\bibfnamefont {T.~A.}\ \bibnamefont {Maier}}, \bibinfo {author} {\bibfnamefont {M.~S.}\ \bibnamefont {Jarrell}},\ and\ \bibinfo {author} {\bibfnamefont {D.~J.}\ \bibnamefont {Scalapino}},\ }\bibfield  {title} {\bibinfo {title} {Structure of the pairing interaction in the two-dimensional hubbard model},\ }\href {https://doi.org/10.1103/PhysRevLett.96.047005} {\bibfield  {journal} {\bibinfo  {journal} {Phys. Rev. Lett.}\ }\textbf {\bibinfo {volume} {96}},\ \bibinfo {pages} {047005} (\bibinfo {year} {2006})}\BibitemShut {NoStop}%
\bibitem [{\citenamefont {Scalapino}(2007)}]{scalapino2007numerical}%
  \BibitemOpen
  \bibfield  {author} {\bibinfo {author} {\bibfnamefont {D.~J.}\ \bibnamefont {Scalapino}},\ }\href {https://doi.org/10.1007/978-0-387-68734-6_13} {\emph {\bibinfo {title} {Handbook of High-Temperature Superconductivity: Theory and Experiment}}}\ (\bibinfo  {publisher} {Springer},\ \bibinfo {address} {New York, NY},\ \bibinfo {year} {2007})\ pp.\ \bibinfo {pages} {495--526}\BibitemShut {NoStop}%
\bibitem [{\citenamefont {Qin}\ \emph {et~al.}(2022)\citenamefont {Qin}, \citenamefont {Schäfer}, \citenamefont {Andergassen}, \citenamefont {Corboz},\ and\ \citenamefont {Gull}}]{Hubreview1}%
  \BibitemOpen
  \bibfield  {author} {\bibinfo {author} {\bibfnamefont {M.}~\bibnamefont {Qin}}, \bibinfo {author} {\bibfnamefont {T.}~\bibnamefont {Schäfer}}, \bibinfo {author} {\bibfnamefont {S.}~\bibnamefont {Andergassen}}, \bibinfo {author} {\bibfnamefont {P.}~\bibnamefont {Corboz}},\ and\ \bibinfo {author} {\bibfnamefont {E.}~\bibnamefont {Gull}},\ }\bibfield  {title} {\bibinfo {title} {The hubbard model: A computational perspective},\ }\href {https://doi.org/https://doi.org/10.1146/annurev-conmatphys-090921-033948} {\bibfield  {journal} {\bibinfo  {journal} {Annual Review of Condensed Matter Physics}\ }\textbf {\bibinfo {volume} {13}},\ \bibinfo {pages} {275} (\bibinfo {year} {2022})}\BibitemShut {NoStop}%
\bibitem [{\citenamefont {Maier}\ and\ \citenamefont {Scalapino}(2019)}]{Maier2019}%
  \BibitemOpen
  \bibfield  {author} {\bibinfo {author} {\bibfnamefont {T.~A.}\ \bibnamefont {Maier}}\ and\ \bibinfo {author} {\bibfnamefont {D.~J.}\ \bibnamefont {Scalapino}},\ }\bibfield  {title} {\bibinfo {title} {Pairfield fluctuations of a 2d hubbard model},\ }\href {https://doi.org/10.1038/s41535-019-0169-9} {\bibfield  {journal} {\bibinfo  {journal} {npj Quantum Materials}\ }\textbf {\bibinfo {volume} {4}},\ \bibinfo {pages} {1} (\bibinfo {year} {2019})}\BibitemShut {NoStop}%
\bibitem [{\citenamefont {Maier}\ \emph {et~al.}(2008)\citenamefont {Maier}, \citenamefont {Poilblanc},\ and\ \citenamefont {Scalapino}}]{PhysRevLett.100.237001}%
  \BibitemOpen
  \bibfield  {author} {\bibinfo {author} {\bibfnamefont {T.~A.}\ \bibnamefont {Maier}}, \bibinfo {author} {\bibfnamefont {D.}~\bibnamefont {Poilblanc}},\ and\ \bibinfo {author} {\bibfnamefont {D.~J.}\ \bibnamefont {Scalapino}},\ }\bibfield  {title} {\bibinfo {title} {Dynamics of the pairing interaction in the hubbard and $t-j$ models of high-temperature superconductors},\ }\href {https://doi.org/10.1103/PhysRevLett.100.237001} {\bibfield  {journal} {\bibinfo  {journal} {Phys. Rev. Lett.}\ }\textbf {\bibinfo {volume} {100}},\ \bibinfo {pages} {237001} (\bibinfo {year} {2008})}\BibitemShut {NoStop}%
\bibitem [{\citenamefont {Jiang}\ \emph {et~al.}(2018)\citenamefont {Jiang}, \citenamefont {H\"ahner}, \citenamefont {Schulthess},\ and\ \citenamefont {Maier}}]{Mi2018}%
  \BibitemOpen
  \bibfield  {author} {\bibinfo {author} {\bibfnamefont {M.}~\bibnamefont {Jiang}}, \bibinfo {author} {\bibfnamefont {U.~R.}\ \bibnamefont {H\"ahner}}, \bibinfo {author} {\bibfnamefont {T.~C.}\ \bibnamefont {Schulthess}},\ and\ \bibinfo {author} {\bibfnamefont {T.~A.}\ \bibnamefont {Maier}},\ }\bibfield  {title} {\bibinfo {title} {$d$-wave superconductivity in the presence of nearest-neighbor coulomb repulsion},\ }\href {https://doi.org/10.1103/PhysRevB.97.184507} {\bibfield  {journal} {\bibinfo  {journal} {Phys. Rev. B}\ }\textbf {\bibinfo {volume} {97}},\ \bibinfo {pages} {184507} (\bibinfo {year} {2018})}\BibitemShut {NoStop}%
\bibitem [{\citenamefont {Zheng}\ and\ \citenamefont {W\'u}(2025)}]{yaoyuan}%
  \BibitemOpen
  \bibfield  {author} {\bibinfo {author} {\bibfnamefont {Y.-Y.}\ \bibnamefont {Zheng}}\ and\ \bibinfo {author} {\bibfnamefont {W.}~\bibnamefont {W\'u}},\ }\bibfield  {title} {\bibinfo {title} {${s}_{\ifmmode\pm\else\textpm\fi{}}$-wave superconductivity in the bilayer two-orbital hubbard model},\ }\href {https://doi.org/10.1103/PhysRevB.111.035108} {\bibfield  {journal} {\bibinfo  {journal} {Phys. Rev. B}\ }\textbf {\bibinfo {volume} {111}},\ \bibinfo {pages} {035108} (\bibinfo {year} {2025})}\BibitemShut {NoStop}%
\end{thebibliography}%
\end{document}